\documentclass[showpacs,aps,amssymb,pra]{revtex4}

\usepackage{bm}
\usepackage{epsfig}
\usepackage{amsmath} 
\usepackage{float}
\usepackage{verbatim}

\def\bv{{\bf v}}
\def\calH{\mathcal H}

\def\beq{\begin{equation}}
\def\eeq{\end{equation}}
\def\nn{\nonumber}

\begin{document}
\title {Mean-field yrast spectrum and persistent currents in a 
two-component Bose gas with interaction asymmetry }
\author{Z. Wu$^1$ and E. Zaremba$^2$}
\affiliation{$^1$Department of Physics and Astronomy, Aarhus University, DK-8000 Aarhus C, Denmark
\\$^2$Department of Physics, Engineering Physics and Astronomy, Queen's University,
Kingston, Ontario, K7L 3N6, Canada}
\author{J. Smyrnakis$^3$, M. Magiropoulos$^3$, Nikolaos K. Efremidis$^4$,
and G. M. Kavoulakis$^3$}
\affiliation{$^3$Technological Education Institute of Crete, P.O. 
Box 1939, GR-71004, Heraklion, Greece
\\
$^4$Department of Applied Mathematics, University of Crete, Heraklion 
71409, Greece}
\date{\today}
\begin{abstract}
We analyze the mean-field yrast spectrum of a two-component Bose gas in the 
ring geometry with arbitrary interaction asymmetry. Of particular interest is the possibility that the yrast spectrum
develops local minima at which persistent superfluid flow can occur. By analyzing the mean-field energy functional, 
we show that local minima can be found at plane-wave states and arise when the system parameters satisfy certain inequalities. 
We then go on to show that these plane-wave states can be yrast states even when the yrast spectrum no longer
exhibits a local minimum. Finally, we obtain conditions which establish when the plane-wave states cease to be
yrast states. Specific examples illustrating the roles played by the various interaction asymmetries are presented.
\end{abstract}
\maketitle
\section{Introduction}
The experimental realization of annular trapping potentials~\cite{gupta05,ryu07,sherlock11}
 has recently led to the observation of persistent superfluid  flow
in a multiply-connected geometry~\cite{ryu07,ramanathan11,moulder12,beattie13}. 
The question of the stability of these superfluid currents is a complex matter and depends
on the nature of the dynamical excitations available to the system. The current consensus is that stability is limited
by the penetration of vortices through the edge of the superfluid with a concomitant change in the phase of the superfluid
order parameter. Several theoretical studies support this scenario~\cite{anglin01,dubessy12,yakimenko13,
abad14,yakimenko15}.

However, underlying these dynamical instabilities is the inherent metastability of the superfluid system. As 
emphasized by Bloch~\cite{Bloch}, this metastability is revealed through the energy of the superfluid as a function 
of its angular momentum, its so-called yrast spectrum~\cite{Mottelson}.
In this paper, we investigate the yrast spectrum of a two-component Bose gas in the ring
geometry. Specifically, we have in mind the situation in which the atoms are confined to a torus where the transverse confinement
is so tight that the system is effectively one-dimensional. When the two species have equal masses $M$, it can be shown quite generally~\cite{Smyrnakis1,Anoshkin} that 
the yrast spectrum takes the form
\beq
E_0(L) = \frac{L^2}{2M_TR^2} +e_0(L),
\eeq
where $R$ is the radius of the ring, $L$ is the total angular momentum and $M_T=MN$ is the 
total mass of the system. Here, $N=N_A+N_B$ is the total number of atoms of type $A$ and $B$.
The function $e_0(L)$ has inversion symmetry and possesses the periodicity property
$e_0(L+N\hbar)=e_0(L).$ 

\begin{center}
\begin{figure}[ht]
     \includegraphics[width=0.49\linewidth]{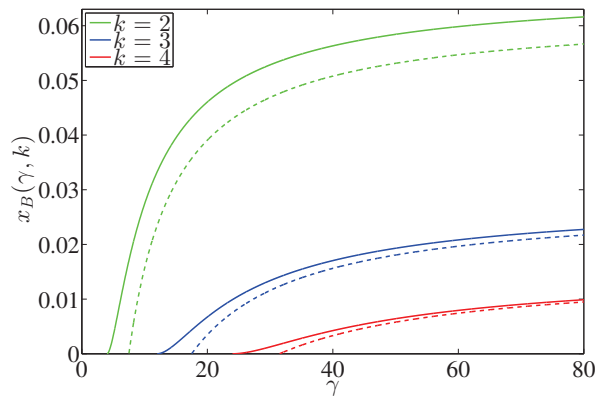}
\caption{The solid lines are critical curves for the emergence of
plane-wave yrast states. In the region bounded by $x_{B}(\gamma,k)$ and $x_{B}(\gamma,k+1)$, 
the plane-wave states $( \phi_0,\phi_q)$, with $q=1,...,k$ are yrast states. The critical curve for $k=1$,
$x_B(\gamma,1) = 0.5$, is not shown. The dashed lines are critical 
curves defining the regions in which the  $( \phi_k,\phi_0)$ state supports persistent currents.} 
\label{gammaxb}
\end{figure}
\end{center}

The above properties of the yrast spectrum are independent of the detailed nature of the inter-particle interactions~\cite{Anoshkin}.
For contact interactions, the interactions can be characterised by the dimensionless parameters $\gamma_{ss'}$ where the subscripts
$s$ and $s'$ take on the values $A$ and $B$. (A detailed definition of these interaction parameters is given in Sec. II.)
However, as shown in several previous 
studies~\cite{Smyrnakis1,Smyrnakis2,Smyrnakis3, Anoshkin,Wu}, the 
{\it mean-field} yrast spectrum of the two-component system, with the added restriction that all
interaction strengths have a common value $\gamma$,
exhibits two additional properties. First, the part of the spectrum in the fundamental
range $0\le l\le 1/2 $, where $l = L/N\hbar$, is not  in general an analytic function 
of the (dimensionless) angular momentum per particle. In particular, the derivative of the 
spectrum is found to exhibit discontinuities at $l=qx_B$, where $x_B = N_B/N$ is the minority 
concentration and $q = 1,2,..., k$. The number of discontinuities $k$
depends on the two relevant parameters of the model, namely $\gamma$ and $x_B$. More 
specifically, it was established that $k$ derivative discontinuities 
occur when the coordinate $(\gamma, x_B)$ lies within a region 
bounded by the two critical curves $x_{B}(\gamma,k)$ and $x_{B}(\gamma,k+1)$ 
in the $\gamma$-$x_B$ plane~\cite{Wu}. These curves are illustrated by the 
solid lines in Fig.~\ref{gammaxb} for $k = 
2,...,4$. Importantly, the point of 
non-analyticity, $l = q x_B$, is associated with the condensate wavefunctions 
having a plane-wave form, $(\psi_A,\psi_B)= ( \phi_0,\phi_q)$, where 
$\phi_q=e^{i q\theta}/\sqrt{2\pi}$. As a result, the relevant 
critical $x_B(\gamma,k)$ curves in Fig.~\ref{gammaxb} can also be viewed as 
defining the regions within which plane-wave yrast states emerge. For values of $l$ other than
these special values, the yrast state is in general a soliton state.

Now, because of the periodicity and inversion symmetry of $e_0(L)$,
a derivative discontinuity at $l = kx_B$ implies discontinuities at 
$l=\mu \pm kx_B$ as well, where $\mu$ is any non-zero integer. The yrast 
states at these angular momenta are $(\psi_A,\psi_B)= ( 
\phi_\mu,\phi_{\mu\pm k})$. The second important property of the yrast spectrum 
concerns a subset of such non-analytic points, namely those at 
$l=k-kx_B = kx_A$. It can be shown~\cite{Wu} that the yrast 
spectrum has local minima at these angular momenta (and only these 
from all possible $l=\mu \pm kx_B$)  for any integer $k$ provided 
$\gamma$ exceeds the critical interaction strength 
\beq
\gamma_{{\rm cr},k} = \frac{4k^2-1}{2(1-4x_Bk^2)}.
\eeq
This expression provides another set of critical 
$x_{B}(\gamma,k)$ curves, which are indicated by the dashed lines in 
Fig.~\ref{gammaxb}. Since a local minimum at $l=k-kx_B$ necessarily 
implies that the corresponding state $(\psi_A,\psi_B)= ( \phi_k,\phi_0)$ 
is already an yrast state, $\gamma > \gamma_{{\rm cr},k}$ 
is thus a sufficient, but not necessary, condition for the existence of plane-wave 
yrast states. This is reflected in the fact that the dashed curves in Fig.~\ref{gammaxb} are 
displaced to the right of the solid curves; with increasing $\gamma$ for a fixed $x_B$, one first
crosses a solid curve at which point some plane-wave state becomes an yrast state, and then 
the dashed curve beyond which this plane-wave state is a local minimum.
The existence of an energy
minimum is of particular significance since, as argued by Bloch~\cite{Bloch}, 
it implies the possibility of persistent superfluid flow. Thus, the 
condition $\gamma > \gamma_{{\rm cr},k}$ can be taken as the stability condition for 
persistent currents at the angular momentum $l=k-k x_B$.

The above conclusions, pertaining to a system with symmetrical 
interaction strengths, were reached with aid of analytic soliton 
solutions to the coupled Gross-Pitaevskii equations, from which the full 
yrast spectrum could be determined~\cite{Wu}. These conclusions were
confirmed by Smyrnakis {\it et al.}~\cite{Smyrnakis14} using an alternative approach.
Since the symmetrical model is rather 
special, it is unclear whether the aforementioned properties of the mean-field
yrast spectrum remain valid for the asymmetrical model in which
the interparticle interactions take on different values. 
This is the main question to be addressed in this paper. To answer it we adopt the
strategy, motivated by the symmetrical model, of examining the mean-field energy functional
in the vicinity of plane-wave states. Even though analytic soliton solutions are not known for the 
asymmetrical model, we are able to use a perturbative analysis to determine the general
behaviour of the energy functional near plane-wave states and obtain
critical conditions analogous to those displayed in Fig.~\ref{gammaxb}. Since all of the 
results for the symmetrical model are recovered by this approach, we are confident
that the conditions we derive do in fact determine the stability of persistent currents in the 
asymmetrical model.

The rest of the paper is organized as follows.  In Sec.~\ref{Stability}, we derive inequalities
involving the system parameters
which establish whether a given plane-wave state is a local minimum of the Gross-Pitaevskii
energy functional. Such a state has a specific angular momentum $l$. We then argue that 
the lowest-energy plane-wave state having this angular momentum is a global minimum if the
inequalities for this state are satisfied and, hence, is an yrast state. These predictions are then checked against known
limiting situations, including that of the symmetric model. In Sec.~\ref{critical_conditions} we then analyze in more detail
the behaviour of the yrast spectrum in the vicinity of a plane-wave state corresponding to a global minimum. 
We first establish that the yrast spectrum has a derivative discontinuity at the angular momentum $l$ of this
state. The instability of persistent currents at this angular momentum is then signalled by the critical condition that one of the
slopes of the yrast spectrum vanishes. At this point, the plane-wave state is no longer a local minimum.
We then obtain a subsequent critical condition for the disappearance of the derivative discontinuity. This condition provides a bound
for the plane-wave state to be an yrast state. We conclude this section with some applications of 
these critical conditions in the determination of plane-wave yrast states. 
The main results of the paper are summarized in the final section. 
 
\section{Local minima of the energy functional at plane-wave states}
\label{Stability}
The system of interest is a two-species Bose gas consisting of 
$N_A$ particles of type $A$ and $N_B$ particles of
type $B$ confined to a ring of radius $R$.  We assume that 
the two species have the same mass $M$. 
Within a mean-field description, the condensate wave
functions $\psi_A$ and $\psi_B$ define the Gross-Pitaevskii (GP) energy
functional (in units of $N \hbar^2 /2MR^2$)
\begin{align}
\bar E [\psi_A,\psi_B] =& \int_0^{2\pi}d\theta \left ( x_A\left 
|\frac{d\psi_A}{d\theta} \right |^2+x_B\left |\frac{d\psi_B}{d\theta} 
\right |^2\right ) +x_A^2\pi \gamma_{AA}\int_0^{2\pi}d\theta 
|\psi_A(\theta)|^4+x_B^2\pi \gamma_{BB}\int_0^{2\pi}d\theta 
|\psi_B(\theta)|^4 \nn \\
&+2x_Ax_B \pi \gamma_{AB}\int_0^{2\pi}d\theta |\psi_A(\theta)|^2 
|\psi_B(\theta)|^2,
\label{Efunc}
\end{align}
where the dimensionless interaction parameters are defined as
$\gamma_{ss'}=U_{ss'}NMR^2/\pi\hbar^2$ with $N= N_A+N_B$.
For the most part, we are concerned with repulsive interactions, $\gamma_{ss'} > 0$.
With the condensate wave functions normalized according to
\beq
\int_0^{2\pi} d\theta |\psi_s(\theta)|^2 = 1,
\label{normalization}
\eeq
the energy per particle defined in Eq.~(\ref{Efunc}) depends on the
particle numbers only through the concentrations $x_s =
N_s/N$.

Our ultimate objective is the determination of the yrast
spectrum which is defined by the lowest energy of the system as a function of the angular
momentum. In units of $N\hbar$, the total angular momentum of
the system is given by
\beq
\bar L[\psi_A,\psi_B] = \sum_s \frac{x_s}{i} \int_0^{2\pi} 
d\theta \psi_s^*(\theta) \frac{\partial}{\partial \theta} 
\psi_s(\theta).
\label{ang_mom}
\eeq
The minimization of Eq.~(\ref{Efunc}) with the constraint $\bar
L[\psi_A,\psi_B] = l$ gives the yrast energy
\beq
\bar E_0(l) = l^2 + e_0(l),
\eeq
where $e_0(l)$ is an even function of $l$ with the periodicity
property $e_0(l+n) = e_0(l)$, $n$ being an arbitrary integer~\cite{Anoshkin}.
As a result of these properties, the yrast spectrum is completely determined
by the behaviour of $e_0(l)$ in the interval $0\le l \le 1/2$. If a point $l_0$ in this 
interval is a point of nonanalyticity of $e_0(l)$, then the points $n\pm l_0$ for
any integer $n$ are points of nonanalyticity of $\bar E_0(l)$. As we shall show,
these points occur at plane-wave yrast states.

We are therefore led to an investigation of the behaviour of the GP energy
functional in the vicinity of some arbitrary
plane-wave state $(\phi_\mu,\phi_{\nu})$.
As discussed in the Introduction, the conditions for which such a state is
an yrast state are known in the case of the symmectrical model. There it
is found that the yrast spectrum exhibits a derivative
discontinuity at the angular momentum corresponding to this
state, namely at
\beq
l= \mu x_A + \nu x_B \equiv \mu + k x_B,
\eeq
where $k = \nu -\mu$. Furthermore, the criteria for the yrast
spectrum exhibiting a local minimum at one of these angular
momenta is also known.
The question we wish to address in this
paper is the extent to which such states can be yrast states
in the asymmetrical model. 

For the $(\phi_\mu, \phi_\nu)$ plane-wave state we have
\beq
\bar E[\phi_\mu,\phi_\nu] = \bar E_{\rm int} + l^2 + x_A x_B k^2,
\label{E_pw}
\eeq
with
\beq
\bar E_{\rm int} = \frac{1}{2} (x_A^2 \gamma_{AA} + 
2x_A x_B \gamma_{AB} + x_B^2 \gamma_{BB}).
\eeq
All such plane-wave states have the same interaction energy
$\bar E_{\rm int}$ but a kinetic energy which depends on
the parameters $\mu$ and $\nu$ or alternatively, $l$ and $k$. 

In searching for the minimum energy plane-wave state of a given
angular momentum $l$, it is useful to note that $x_B = N_B/N$ is in
general a rational number which we denote by
\beq
x_B = \frac{p}{q}
\eeq
where $p$ and $q$ are positive integers having no common divisor. One can easily check that
the plane-wave states ($\phi_{\mu'},\phi_{\mu'+k'}$) defined by the parameters
\beq
\mu'= \mu + mp, \quad k' = k-mq,
\eeq
where $m = 0,\,\pm 1, \,\pm 2,...$, all have the same angular
momentum $l$. In view of Eq.~(\ref{E_pw}), the lowest energy state from 
this infinite set is obtained for the smallest value of $|k'|$. 
This value of  $k'$ will be found in the range
\beq
-\left [ \frac{q}{2} \right ] \le k' \le \left [\frac{q}{2}\right ]
\label{k-range}
\eeq
where $[q/2]$ is the largest integer less than or equal to $q/2$, 
that is, the floor of $q/2$. 
 It is worth noting that the allowed angular momentum values of the plane-waves states take the form 
$l = Q/q$, where $Q$ is an arbitrary integer. Of course, when the restriction
$0\le l\le 1/2$ is imposed, one need only consider $Q=0,1,...,[q/2]$.

Let us now consider the plane-wave state ($\phi_\mu,\phi_{\mu+k}$) with $k$ restricted to 
the range given by Eq.~(\ref{k-range}).
If $q$ is odd, the set of $k$ 
values in this range corresponds to a complete
residue system modulo $q$. Thus, when each possible value of $k$ in
this range is paired with each possible value 
of $\mu$, all possible values of the angular momentum $l$ are generated
without duplication.
As a result, the value of $k$ which minimizes the energy for 
a given $l$ is unique.
The situation for even $q$ is slightly different since $-[q/2]$
and $[q/2]$ are congruent and there are two plane-wave states,
namely ($\phi_\mu$, $\phi_{\mu+q/2}$) and ($\phi_{\mu+p}$, $\phi_{\mu+p-q/2}$), which have the same 
angular momentum and energy. Thus, one cannot decide which of these two states is a potential
yrast state at this particular angular momentum. However, we do
have a prescription for selecting, from all
possible plane-wave states having the same $l$, the specific
state(s) that are potential yrast states.

We note that if $x_B$ is treated as a continuous variable,
it will of course take on irrational values. In this case, no
two plane-wave states will have the same angular momentum and
the complexities associated with rational $x_B$ are avoided.
However, whenever $x_B$ takes on a rational value, the
above considerations will once again apply.


Although our analysis could be restricted to
the plane-wave states with $k$ in the range specified by  Eq.~(\ref{k-range}), 
it is more convenient to consider in the following the GP energy functional
in the vicinity of an arbitrary $(\phi_\mu,\phi_\nu)$ plane-wave state.
To be specific, our goal is to establish
the conditions for which the energy functional will exhibit a local minimum
at this state. To this end, we consider
states $(\psi_A,\psi_B)$ which deviate slightly from
$(\phi_\mu,\phi_{\nu})$, {\it viz.}
\begin{equation}
\label{4.2_wf_perturbed}
 \psi_A =\phi_\mu+\delta\psi_A, \hspace{1cm} 
 \psi_B =\phi_\nu+\delta\psi_B,
\end{equation}
where the deviations are expressed in the form
\begin{align}
\label{4.2_wf_deviations}
\delta\psi_A & = \delta c_{\mu}\phi_{\mu }+\sum\limits_{m> 0}\left (\delta 
c_{\mu+m}\phi_{\mu + m}+\delta c_{\mu-m}\phi_{\mu - m}\right ), \\
\delta\psi_B & = \delta d_{\nu}\phi_{\nu }+\sum\limits_{m> 0}\left (\delta 
d_{\nu+m}\phi_{\nu + m}+\delta d_{\nu-m}\phi_{\nu - m}\right ).
\end{align}
The normalization of these states implies
\begin{align}
\label{A_norm_alt}
|1+\delta c_\mu|^2 + \sum_{n\ne \mu} |\delta c_n|^2 & =1 \\
\label{B_norm_alt}
|1+\delta d_\nu|^2 +\sum_{n\ne \nu}|\delta d_n|^2 & =1.
\end{align}
These relations indicate, for example, that $|1+\delta c_\mu|^2 \le 1$
and $\sum_{n\ne \mu} |\delta c_n|^2 \le 1$.
Alternatively, these normalization conditions can be expressed as
\begin{align}
\label{A_norm}
\delta c_\mu +\delta c_\mu^*+\sum_n |\delta c_n|^2 & =0 \\
\label{B_norm}
\delta d_\nu +\delta d_\nu^*+\sum_n |\delta d_n|^2 & =0.
\end{align}

Substituting Eq.~(\ref{4.2_wf_perturbed}) into Eq.~(\ref{Efunc}) and 
eliminating $\delta c_\mu$ and $\delta d_\nu$ by means of
Eqs.~(\ref{A_norm}) and (\ref{B_norm}),
we find that the change in energy to second order in the deviations is given by 
\begin{align}
\delta \bar E[\psi_A,\psi_B] & =\bar E[\psi_A,\psi_B] -\bar 
E[\phi_\mu,\phi_\nu]  \nn \\
&\simeq \sum_{m>0}\bv^\dag_m {\calH}_m \bv_m,
\label{dFF}
\end{align}
where $\bv_m =(\delta c_{\mu-m}\,\, \delta c^*_{\mu +m}\,\, \delta 
d_{\nu-m}\,\, \delta d^*_{\nu+m})^{\rm T}$ and 
\beq
\calH_m=\left (\begin{array}{cccc}
x_A(x_A\gamma_{AA}+m^2-2\mu m) & x_A^2\gamma_{AA} &x_Ax_B\gamma_{AB} & 
x_Ax_B\gamma_{AB} \\
 x_A^2\gamma_{AA}& x_A(x_A\gamma_{AA}+m^2+2\mu m) & x_Ax_B\gamma_{AB} & 
x_Ax_B\gamma_{AB} \\
 x_Ax_B\gamma_{AB}&x_Ax_B\gamma_{AB} &x_B(x_B\gamma_{BB}+m^2-2\nu m) 
&x_B^2\gamma_{BB} \\
x_Ax_B\gamma_{AB}&x_Ax_B\gamma_{AB} &x_B^2\gamma_{BB} 
&x_B(x_B\gamma_{BB}+m^2+2\nu m) 
\end{array} \right ).
\label{H_matrix}
\eeq
We thus see that the change in energy is a quadtratic form.

If the matrices $\calH_m$ are all positive definite, the
energy $\bar E[\phi_\mu,\phi_\nu]$ is a local minimum in the
function space defined by $\psi_A$ and $\psi_B$. According to
Sylvester's criterion~\cite{Gelfand}, positive-definiteness is assured if all 
the leading principal minors of $\calH_m$ are positive, namely
\begin{align}
\label{ineq1}
x_A\gamma_{AA}+m^2-2\mu m > 0, \\
\label{ineq2}
2x_A\gamma_{AA} +m^2 -4\mu^2 >0, \\
\label{ineq3}
\left (2x_A\gamma_{AA}+m^2-4\mu^2\right )\left (x_B\gamma_{BB}+m^2-2\nu m\right ) 
-2x_Ax_B\gamma_{AB}^2>0, \\
\label{ineq4}
\left (2x_A\gamma_{AA}+m^2-4\mu^2 \right ) \left ( 
2x_B\gamma_{BB}+m^2-4\nu^2\right ) - 4x_Ax_B\gamma^2_{AB}>0.
\end{align}
It is straighforward to show that Eq.~(\ref{ineq2}) implies
Eq.~(\ref{ineq1}); likewise, Eq.~(\ref{ineq4}) together with
Eq.~(\ref{ineq2}) implies Eq.~(\ref{ineq3}). Thus, Eqs.~(\ref{ineq2}) and
(\ref{ineq4}) are the fundamental inequalities determining the
positive-definiteness of ${\cal H}_m$. Furthermore, these 
inequalities are satisfied for all $m$ if they are satisfied 
for $m=1$. We thus see that the inequalities
\begin{align}
\label{stab1}
2x_A\gamma_{AA} +1 -4\mu^2 >0, \\
\left (2x_A\gamma_{AA}+1-4\mu^2 \right ) \left ( 
2x_B\gamma_{BB}+1-4\nu^2\right ) - 4x_Ax_B\gamma^2_{AB}>0
\label{stab2}
\end{align}
are the necessary and sufficient conditions for
$(\phi_\mu,\,\phi_\nu)$ being a local minimum in the function
space. It is important to note that, although the state
$(\phi_\mu,\,\phi_\nu)$ has the angular momentum
$l = \mu x_A+\nu x_B$, the variations in Eq.~(\ref{4.2_wf_perturbed}) 
allow for deviations of the angular momentum from this value. In other words, the local
minimum that we are finding is not constrained by the angular momentum $l$;
the local minimum exists for arbitrary variations of the condensate wave functions about
the plane-wave state of interest.

On the other hand, it is possible to consider variations which are 
further constrained (apart from normalization) by the angular
momentum $l = \mu x_A+\nu x_B$; such states define a hypersurface in
function space. The state $(\phi_\mu,\,\phi_\nu)$ lies on this
surface and, if the inequalities in Eqs.~(\ref{stab1}) and
(\ref{stab2}) are satisfied, its energy is lower than that of any other state 
in its vicinity on the hypersurface. If this state
were in fact a {\it global} minimum on the hypersurface, it 
would be, by definition, an yrast state. Since the specific plane-wave state $(\phi_\mu,\phi_{\nu})$, where $\nu = \mu +k$ with $k$
restricted to the range in Eq.~(\ref{k-range}), has the lowest energy of all the plane-wave
states having the same angular momentum, it is clearly a candidate for being
the global minimum. In the Appendix,
we show that conditions exist for which such a state 
 is assured to
be a global minimum on the $l=\mu+kx_B$ hypersurface.  We now make the stronger assumption
that this specific plane-wave state is a global minimum
when the inequalities
in Eqs.~(\ref{stab1}) and (\ref{stab2}) are satisfied, and is hence 
an yrast state. As we shall show, this assumption is consistent with the 
results of the symmetric model and, for reasons of continuity, 
would be expected to continue holding
as the interaction parameters gradually become
asymmetrical. Furthermore, 
the inequalities in Eqs.~(\ref{stab1}) and (\ref{stab2})
also ensure that the energy increases as one moves away from  $(\phi_\mu,\phi_{\nu})$
in directions of either increasing or decreasing angular momenta. 
According to the Bloch criterion, this would imply that
persistent currents are stable at the $l=\mu+kx_B$ angular momentum point of the yrast spectrum. 
We now consider some special cases in
order to make contact with earlier work. 
Unless stated otherwise, the $\mu$ and $\nu$ indices will henceforth refer to plane-wave states
for which the difference $k=\nu-\mu$ is restricted to the range in Eq.~(\ref{k-range}).

\subsection{Case 1: $\mu = \nu = n$}
This case corresponds to integral angular momenta, $l = n$.
Eq.~(\ref{stab2}) then reduces to 
\beq
\left (x_A\gamma_{AA}-\frac{4n^2-1}{2} \right ) 
\left (x_B\gamma_{BB}-\frac{4n^2-1}{2} \right ) 
> x_Ax_B\gamma^2_{AB}.
\label{integral-l1}
\eeq
This together with Eq.~(\ref{stab1}) implies
\beq
x_A\gamma_{AA} + x_B\gamma_{BB} > 4n^2-1.
\label{integral-l2}
\eeq
These are the two inequalities given in Ref.~\cite{Anoshkin}
that establish the stability of persistent currents at integral 
values of $l$.

For $n=0$, Eq.~(\ref{integral-l1}) reduces to 
\beq
\label{energetic_stability}
\left (x_A\gamma_{AA}+\frac{1}{2} \right ) \left ( 
x_B\gamma_{BB}+\frac{1}{2}\right ) > x_Ax_B\gamma^2_{AB}
\eeq
This is the condition for energetic or dynamic stability and
ensures that the uniform state $(\phi_0,\phi_0)$ is stable against phase
separation. This state gives the absolute
minimum of the GP energy functional, and by virtue of the
periodicity of $e_0(l)$, the states
$(\phi_n,\phi_n)$ with integral angular momentum $l = n$ are {\it
all} yrast states. It is thus clear from a consideration of this special case that the
inequalities in Eqs.~(\ref{stab1}) and (\ref{stab2}) do
in fact define a global minimum on the $l=n$ hypersurface.

If we now consider the special case $\gamma_{AA}\gamma_{BB} = \gamma_{AB}^2$, the inequality in
Eq.~(\ref{integral-l1}) for $n\ne 0$ reduces to
\beq
x_A\gamma_{AA}+x_B\gamma_{BB} < \frac{1}{2} (4n^2-1).
\eeq
This inequality is incompatible with Eq.~(\ref{integral-l2}) 
which implies that the $(\phi_n,\phi_n)$ state
{\it cannot} be a local minimum for these interaction parameters. However,
if Eq.~(\ref{energetic_stability}) is satisfied, this state
is still an yrast state. Thus, a local minimum in function space is 
not a necessary condition for the plane-wave state
being an yrast state. In the following section we will show that
the absence of a local minimum in function space also implies
the lack of a local minimum in the yrast spectrum and
the absence of persistent currents. As explained in
Ref.~\cite{Anoshkin}, the physical reason for the absence of persistent currents
when $\gamma_{AA}\gamma_{BB} = \gamma_{AB}^2$ is that the
Bogoliubov excitations exhibit a particle-like dispersion which
destabilizes superfluid flow.

\subsection{Case 2: $\mu \ne \nu$; $\gamma_{AA}\gamma_{BB} = \gamma_{AB}^2$}
In this case, the inequality in Eq.~(\ref{stab2}) reduces to
\beq
(1-4\nu^2)(2x_A\gamma_{AA} + 1 - 4\mu^2)
+2x_B\gamma_{BB}(1-4\mu^2) > 0.
\label{mu.ne.nu}
\eeq
If $\mu = 0$, this inequality implies
\beq
\nu^2 < \frac{1+2x_A \gamma_{AA}+2x_B
\gamma_{BB}}{4(1+2x_A\gamma_{AA})}.
\label{nu-ineq}
\eeq
The values of $\nu$ satisfying this inequality depend on the
values of the parameters $x_A$, $\gamma_{AA}$ and $\gamma_{BB}$.
If $\mu \ge 1$, we have $(1-4\mu^2) < 0$. Thus in view of
Eq.~(\ref{stab1}), the inequality in Eq.~(\ref{mu.ne.nu}) 
can only be satisfied if $\nu = 0$. We
have thus established that local minima can occur at the states
$(\phi_\mu, \phi_0)$ with angular momenta $l = \mu x_A$ or at
$(\phi_0,\phi_\nu)$ with angular momenta $l= \nu x_B$. In this
latter case, however, the range of $\nu$ is limited by the
inequality in Eq.~(\ref{nu-ineq}).

In the symmetric model with $\gamma_{AA} = \gamma_{BB} =
\gamma_{AB} = \gamma$, the only possible value of $\nu$ in
Eq.~(\ref{nu-ineq}) is zero if we take $A$ to be the majority component ($x_A >1/2$). Thus local
minima can only occur for the $(\phi_\mu, \phi_0)$ states in
this case. Furthermore, Eq.~(\ref{mu.ne.nu}) gives
\beq
\gamma > \frac{4\mu^2-1}{2(1-4x_B \mu^2)}.
\eeq
This is precisely the condition for
persistent currents to occur at $l = \mu x_A$ found in
Ref.~\cite{Wu} using the explicit soliton solutions.
Once again we see that the inequalities in Eqs.~(\ref{stab1})
and (\ref{stab2}) predict the stability of persistent currents
at an {\it yrast} state.

Finally, we wish to point out that $\nu$ in Eq.~(\ref{nu-ineq}),
if unrestricted by the range of $k$, 
can be made arbitrarily large by making $\gamma_{AA}$
sufficiently small and $\gamma_{BB}$ sufficiently large. If $x_B
= p/q$, the angular momentum of the $(\phi_0,\phi_\nu)$ state is
$l=\nu p/q$. Choosing $\nu = q$ we have $l = p$ which is also
the angular momentum of the $(\phi_p,\phi_p)$ state. By
Eq.~(\ref{E_pw}), this state has a lower energy than the
$(\phi_0,\phi_q)$ state which therefore cannot be an yrast
state. Thus a local minimum in function space does not mean 
that one is necessarily dealing with an yrast state. However,
as stated earlier, the plane-wave state with the lowest 
energy is a potential yrast state. All the examples we
have considered so far support the assumption that such a state is
indeed an yrast state if the inequalities in Eqs.~(\ref{stab1}) 
and (\ref{stab2}) are satisfied.

\subsection{Arbitrary parameters}
We now make some observations regarding the inequalities in Eqs.~(\ref{stab1}) and (\ref{stab2})
for general values of the parameters. To simplify matters, we consider the way in which
the inequalities can be violated through a variation of only a single parameter.

It is easy to see that an increase of $\gamma_{AB}$ or a decrease of either $\gamma_{AA}$ or 
$\gamma_{BB}$ will eventually lead to a violation of the inequalities, with the consequence
that persistent currents are destabilized. The dependence on $x_B$ is more interesting.
The left hand side of Eq.~(\ref{stab2}) can be written as a quadratic function of $x_B$, {\it viz.},
\beq
h(x_B) = ax_B^2 +bx_B +c,
\label{gxb}
\eeq
with
\begin{eqnarray}
\label{ha}
a&=& 4(\gamma^2_{AB}-\gamma_{AA}\gamma_{BB})\\
\label{hb}
b&=& -2\left [ 2(\gamma^2_{AB}-\gamma_{AA}\gamma_{BB})-(1-4\mu^2)\gamma_{BB}+(1-4\nu^2)\gamma_{AA} \right ]\\
\label{hc}
c &=& (1-4\nu^2)(2\gamma_{AA}+1-4\mu^2)
\end{eqnarray}
Whether or not $h(x_B) >0$ depends on the nature and location of the roots of $h(x_B)$. These 
can be analyzed in terms of the descriminant
\beq
\Delta_{x_B} = b^2-4ac.
\eeq
The critical values of $x_B$ are determined by the roots of $h(x_B)$ which will be analyzed for the following three cases: (i)
$\gamma_{AB}^2-\gamma_{AA}\gamma_{BB}=0$; (ii) $\gamma_{AB}^2-\gamma_{AA}\gamma_{BB}> 0$; and 
(iii) $\gamma_{AB}^2-\gamma_{AA}\gamma_{BB}< 0$. 
Case (i) falls under Case 2 discussed above. 

(ii)  $\gamma_{AB}^2-\gamma_{AA}\gamma_{BB}> 0$: 
If $\Delta_{x_B} <0$, $h(x_B)$ has no real root and since $a>0$, $h(x_B) > 0$ for all $x_B$. Thus, the stability of the persistent currents is solely determined by  Eq.~(\ref{stab1}). This implies that persistent currents are stable for
\beq
0< x_B \le \rm{min}\left \{1,\frac{2\gamma_{AA}+1-4\mu^2}{2\gamma_{AA}}\right \}.
\label{xbstab1}
\eeq
This scenario is only possible for $\nu = 0$ 
since $c<0$ for $\nu >0$ (recall Eq.~(\ref{stab1})). If $\Delta_{x_B}>0 $ for $\nu = 0$, $h(x_B)$ has two negative roots if $b>0$ or two positive roots if $b<0$. The former situation again means that the persistent currents are stable for $x_B$ satisfying Eq.~(\ref{xbstab1}). In the latter situation, the range of stability of persistent currents is determined by the location of the two positive roots relative to the range specified by Eq.~(\ref{xbstab1}). If $\Delta_{x_B}>0$ for $\nu > 0$, $h(x_B)$ has a negative and positive root. If the latter lies in the interval $[0, 1]$, persistent currents are stable for values of $x_B$ greater than the positive root and overlapping with the interval defined by Eq.~(\ref{xbstab1}). 

(iii)  $\gamma_{AB}^2-\gamma_{AA}\gamma_{BB}< 0$:
Here, persistent currents cannot occur for any $x_B$ if $\Delta_{x_B}<0$. Since $a<0$, this is only possible when $c<0$, which implies $\nu >0$. If $\Delta_{x_B}>0$ for $\nu > 0$, then again $h(x_B)$ either has two negative roots or two positive roots. Two negative roots would mean that the persistent currents are not possible for any value of $x_B$. In the case of two positive roots, the range of stability is again determined by the location of the roots relative to the range specified by Eq.~(\ref{xbstab1}). For $\nu = 0$, $h(x_B)$ has one negative and one positive root. In this case, there is always some finite $x_B$-interval within which persistent currents are possible.

\begin{center}
\begin{figure}[ht]
     \includegraphics[width=0.49\linewidth]{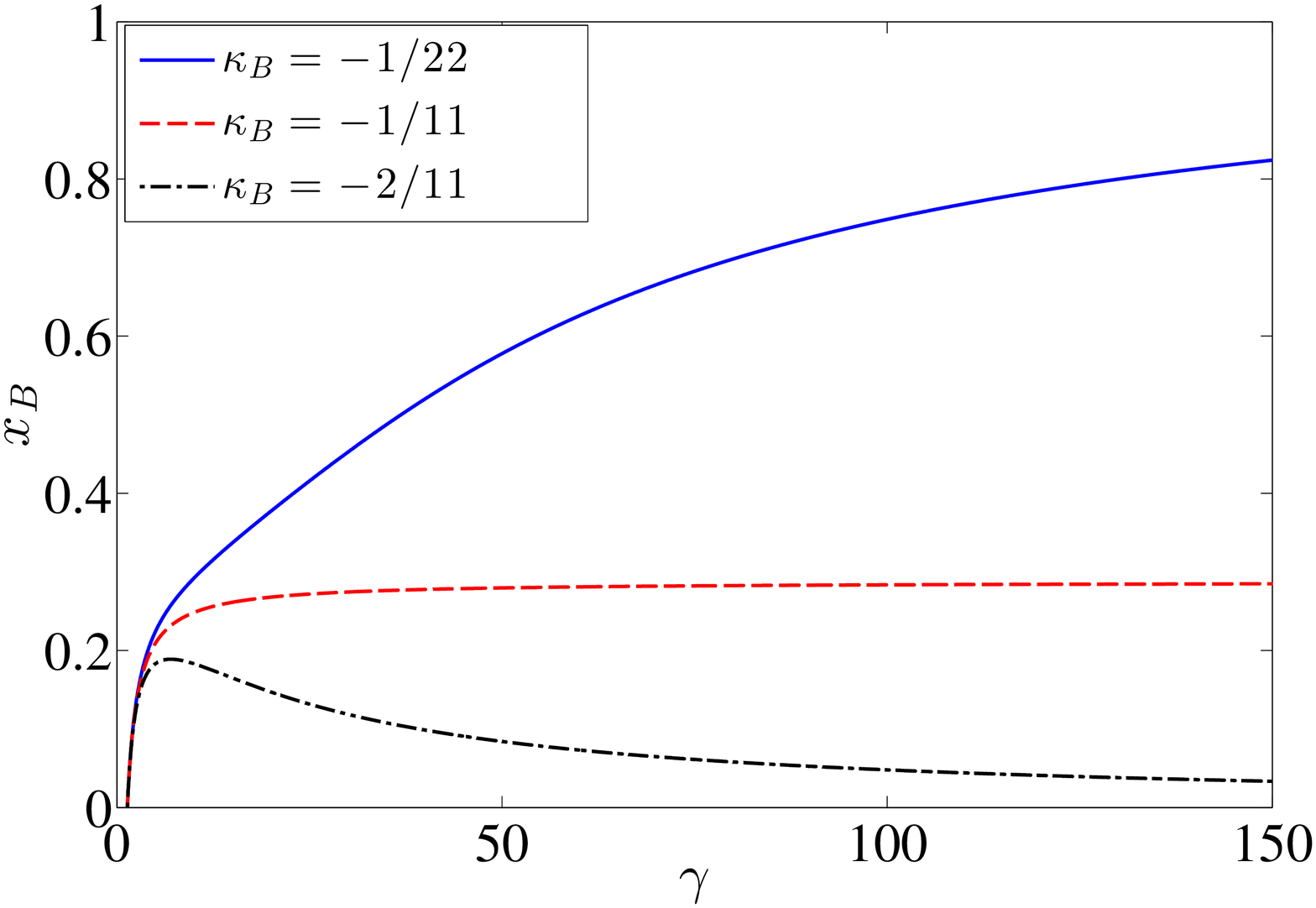}
\caption{The critical $x_B(\gamma)$ curves for persistent currents to be stable at the $(\phi_1,\phi_0)$ state for $\kappa_A = 0.1$ and three different $\kappa_B$ values.} 
\label{pers}
\end{figure}
\end{center}

We now give a simple example illustrating the general discussion given above.
To be specific, we determine the dependence on the interaction asymmetries of the
critical $x_B$ value for which persistent currents are possible at the $(\phi_1,\phi_0)$ plane-wave state.
To facilitate the discussion, we parameterise the interactions as
$\gamma_{AB}=\gamma$, $\gamma_{AA}=(1+\kappa_A)
\gamma$ and $\gamma_{BB}=(1+\kappa_B)\gamma$. The results shown in
Fig.~\ref{pers} are obtained for the fixed value of $\kappa_A = 0.1$ and three $\kappa_B$ values
that are representative of cases (i)-(iii). In all the cases considered here, 
the critical $x_B(\gamma)$ curve is determined by one of the roots of $h(x_B)$.
For $\kappa_B = -1/11$, one finds
$\gamma_{AB}^2-\gamma_{AA}\gamma_{BB} = 0$ and the critical $x_B(\gamma)$ curve
is determined by the only root $x_B(\gamma) = -c/b$, where $b$ and $c$ are specified in 
Eqs.~(\ref{hb})-(\ref{hc}) with the appropriate parameters. 
It is easy to check that the critical curve has an asymptote given by
$x_B = ({1+\kappa_A})/({4+\kappa_A+3\kappa_B})$.
For $\kappa_B = -1/22$, one finds $\gamma_{AB}^2-\gamma_{AA}\gamma_{BB} < 0$
and $h(x_B)$ has one negative and one positive root. The critical curve is given by
the positive root $x_B = (-b+\sqrt{b^2-4ac})/2a$ with an asymptote $x_B=1$.
Finally for $\kappa_B = -2/11$, we have 
$\gamma_{AB}^2-\gamma_{AA}\gamma_{BB}> 0$ and $h(x_B)$ has
two positive roots.  The critical curve is given by
the smaller root $x_B = (-b-\sqrt{b^2-4ac})/2a$ with an asymptote $x_B=0$.
Interestingly, the dependence of $x_B$ on $\gamma$ is not monotonic for $\kappa_B = -1/22$. Thus it
is possible that persistent currents are stable at a fixed value of $x_B$ in only a {\it finite}
interval of $\gamma$. In other words, persistent currents can be stabilized with increasing 
$\gamma$ but are then destabilized with further increases in $\gamma$.

\section{Critical conditions for the existence of plane-wave yrast states}
\label{critical_conditions}
\subsection{General theory}
In the previous section we argued that the plane-wave state $(\phi_\mu,\phi_\nu)$
with the lowest kinetic energy of all plane-wave states having the angular momentum
$l = x_A\mu+x_B\nu$ is an yrast state when this state becomes a local minimum of the
GP energy functional. Furthermore, persistent currents are stable at the angular momentum
corresponding to this plane-wave state. We hypothesized that the validity of these statements 
follows from the inequalities in Eqs.~(\ref{stab1}) and (\ref{stab2}).
In other words, these inequalities
are sufficient conditions for
$( \phi_\mu,\phi_{\nu})$ to be an yrast state, but as already pointed out, they are not
necessary conditions. In this section we investigate the extent to which a necessary condition
can be found. If this
condition is known, it follows from the periodicity
and inversion symmetry of the yrast spectrum that
all the states of the form $
( \phi_n,\phi_{n\pm |k|})$, where $k = \nu - \mu$ and 
$n$ is an arbitrary integer, are yrast states as well.
This observation indicates that the necessary
condition depends on $\mu$ and $\nu$ only through
their absolute difference $|k|$. 

In searching for a necessary condition we require more information about the behaviour of
the yrast spectrum in the vicinity of the plane-wave state.
We first show
that the local energy minimum at
$( \phi_\mu,\phi_{\nu})$ entails a
slope discontinuity of the yrast spectrum at
$l_0=\mu + k x_B$. Thus the condition for the stability of persistent
currents can be expressed in terms of the slopes of the yrast spectrum at
the plane-wave state of interest. For the symmetrical model~\cite{Wu},
one finds that $( \phi_\mu,\phi_{\nu})$
ceases to be an yrast state when the derivative discontinuity
disappears. We cannot state definitively that this is also true for the
asymmetrical model, however, the condition for which the slope discontinuity 
disappears can still be determined. We argue that this condition places a bound on the existence of the
plane-wave yrast state.

To establish the existence of a slope discontinuity, we
investigate the behaviour of the yrast spectrum in the
neighbourhood of the yrast state $( \phi_\mu,\phi_{\nu})$ with angular momentum
$l_0$. For small deviations $\delta l = l - l_0$ of the angular momentum, 
we expect the yrast state at $l$ to deviate only slightly from the plane-wave
state $( \phi_\mu,\phi_{\nu})$ at $l_0$. The yrast state can then be well approximated by
\begin{align}
\label{trwfA}
\psi_A &\simeq \phi_\mu+ \delta c_{\mu-1}\phi_{\mu-1} +
\delta c_\mu \phi_\mu + \delta c_{\mu+1}\phi_{\mu+1} , \\
\psi_B &\simeq \phi_\nu+\delta d_{\nu-1}\phi_{\nu-1} +
\delta d_\nu \phi_\nu + \delta d_{\nu+1}\phi_{\nu+1},
\label{trwfB}
\end{align}
where the coefficients of the deviations are small in absolute
magnitude in comparison to unity and satisfy the
normalization conditions
\begin{align}
\label{normA}
\delta c_\mu +\delta c_\mu^*+|\delta c_{\mu-1}|^2 +
|\delta c_\mu|^2  + |\delta c_{\mu+1}|^2 & =0 \\
\label{normB}
\delta d_\nu +\delta d_\nu^*+ |\delta d_{\nu-1}|^2 + 
|\delta d_\nu|^2  + |\delta d_{\nu+1}|^2 & =0.
\end{align}
Taking these normalization conditions into account, the angular momentum deviation is given by
\beq
\delta l =x_A(|\delta c_{\mu+1}|^2-
|\delta c_{\mu-1}|^2) + x_B(|\delta d_{\nu+1}|^2-|\delta d_{\nu-1}|^2).
\label{dl}
\eeq
This deviation can of course be either positive or negative. We observe that the 
square of the modulus of the coefficients appearing in Eq.~(\ref{dl}) is of order $\delta l$.

The determination of the coefficients is achieved by 
minimizing the energy functional in Eq.~(\ref{Efunc})
with the normalization constraint in Eq.~(\ref{normalization})
and the angular momentum constraint $\bar L[\psi_A,\psi_B] = l = l_0+\delta l$.
To impose this latter constraint, we
introduce a Lagrange multiplier $\Omega$ and
minimize the energy functional
\beq
\bar F[\psi_A,\psi_B] = \bar E[\psi_A,\psi_B]-\Omega\bar 
L[\psi_A,\psi_B].
\label{f_fun}
\eeq
This minimization yields the yrast spectrum $\bar E_0(l)$. The Lagrange
multiplier $\Omega(l)$ obtained in this process is in fact the slope
of the yrast spectrum, namely
\begin{equation}
\Omega(l) = \frac{\partial \bar E_0(l)}{\partial l}.
\label{omega}
\end{equation}
Thus, information about the slope of the yrast spectrum is 
provided by this quantity.

Substituting Eqs.~(\ref{trwfA}) and (\ref{trwfB}) into 
Eq.~(\ref{f_fun}) and
eliminating $\delta c_\mu$ and $\delta d_\nu$ by means of
Eqs.~(\ref{normA}) and (\ref{normB}), we obtain 
\begin{align}
\bar F[\psi_A,\psi_B]\simeq \bar F[\phi_\mu,\phi_\nu] +
\bv_1^\dag {\calH}(\Omega) \bv_1
\label{dF}
\end{align}
to second order in the expansion coefficients.
Here $\bv_1 = (\delta c_{\mu-1}\,\, \delta c^*_{\mu +1}\,\,
\delta d_{\nu-1}\,\, \delta d^*_{\nu+1})^{\rm T}$ and
\beq
\calH(\Omega) = \calH_1 - \Omega \Sigma,
\eeq
where $\calH_1$ is defined by Eq.~(\ref{H_matrix}) with $m=1$ and
$\Sigma$ is the diagonal matrix
\beq
\Sigma=\left (\begin{array}{cccc}
-x_A &0  &0 &0  \\
0 &x_A &0  &0  \\
0 &0 &-x_B &0\\
0&0&0&x_B
\end{array} \right ).
\eeq
In terms of this matrix, the angular momentum deviation is given by
$\delta l = \bv_1^\dag \Sigma \bv_1$.

The extremization of the functional in Eq.~(\ref{dF}) leads to the
following set of linear equations
\beq
\calH(\Omega) \bv_1 = 0.
\label{le}
\eeq
For these equations to have a non-trivial solution,
one must have ${\rm det} (\calH(\Omega)) = 0$. This condition
leads to the equation
\beq
f(\Omega) \equiv \left [(\Omega-2\mu)^2 - 2x_A\gamma_{AA}-1 \right ]
\left [ (\Omega-2\nu)^2-2x_B\gamma_{BB}-1\right ] -
4x_Ax_B\gamma^2_{AB}= 0,
\label{cons1}
\eeq
which is a quartic equation in $\Omega$. We will demonstrate
that two of its solutions in fact correspond to the slopes of 
the yrast spectrum at $l_0 = \mu +kx_B$. 

\begin{center}
\begin{figure}[ht]
     \includegraphics[width=0.85\linewidth]{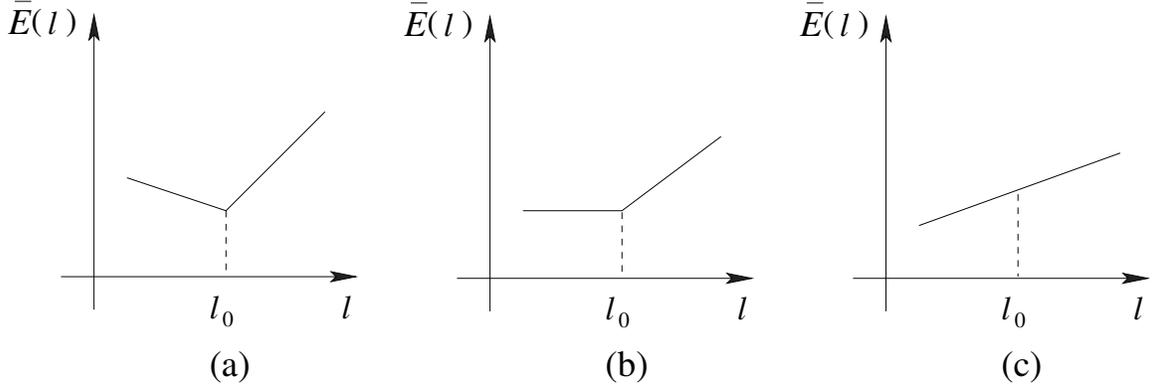}
\caption{Schematic behaviour of the yrast spectrum in the vicinity of a plane-wave yrast state: 
(a) the inequalities in Eqs.~(\ref{stab1}) and (\ref{stab2})
are satisfied and the yrast spectrum shows a cusp-like local minimum; (b) the inequalities are
not satisfied but a slope discontinuity persists; (c) the plane-wave state is no longer an yrast
state and the yrast spectrum is smooth through the angular momentum $l_0$.}
\label{yrast}
\end{figure}
\end{center}

To analyze the zeroes of $f(\Omega)$, we begin by assuming that the inequality in
Eq.~(\ref{stab2}) is satisfied which implies $f(0) > 0$.
We now define the quartic $g(\Omega) \equiv f(\Omega)+4x_A x_B
\gamma_{AB}^2$, which is simply $f(\Omega)$ shifted vertically
by the constant $4x_A x_B\gamma_{AB}^2$. We also have $g(0) >
0$. The solutions of $g(\Omega) = 0$ are
\beq
(2\mu \pm \sqrt{2x_A \gamma_{AA} + 1},\,\,
2\nu \pm \sqrt{2x_B \gamma_{BB} + 1}).
\eeq
If the inequalities in Eqs.~(\ref{stab1}) and
(\ref{stab2}) are satisfied, two of these roots are negative and two are
positive. Furthermore, $g'(0) > 0$,
which implies that $g(\Omega)$ has a maximum at a point $\Omega
> 0$ between the largest negative root and the smallest positive
root. By shifting $g(\Omega)$ down by $4x_A x_B\gamma_{AB}^2$
we recover $f(\Omega)$ and since $f(0) > 0$, we conclude that
$f(\Omega)=0$ must have four distinct, real roots, two of
which are negative and two positive. These roots will be 
denoted by $\Omega_1,...,\Omega_4$ with the ordering
\beq
\Omega_1 < \Omega_2 < 0 < \Omega_3 < \Omega_4.
\eeq
Recalling Eq.~(\ref{omega}), these values are to be identified with the 
slope of the yrast spectrum at $l = l_0$. 
The fact that the energy at $l_0$ is a local minimum when the
inequalities in Eqs.~(\ref{stab1}) and (\ref{stab2})
are satisfied
now implies that the left and right slopes of the yrast spectrum
are necessarily $\Omega_2$ and $\Omega_3$, respectively.
This establishes the fact that the slope of $\bar E_0(l)$
at $l _0$
is discontinuous if $\bar E[\psi_A,\psi_B]$ has a local minimum 
at $(\phi_\mu,\phi_\nu)$. The qualitative behaviour of the yrast spectrum
in the vicinity of $l_0$ is shown in Fig.~\ref{yrast}(a).

Once the roots of Eq.~(\ref{cons1}) have been determined,
Eq.~(\ref{le}) can be solved to yield
\begin{align}
\frac{\delta c^*_{\mu+1}}{\delta c_{\mu-1}} &= \frac{-2\mu+1
+\Omega }{2\mu+1-\Omega }; \nn \\
\frac{\delta d_{\nu-1}}{\delta c_{\mu-1}}&=\frac{(2\nu+1-\Omega)
\left [(\Omega-2\mu)^2-2x_A\gamma_{AA}-1\right ]}{2x_B\gamma_{AB}
(2\mu+1-\Omega)}; \nn \\
\frac{\delta d^*_{\nu+1}}{\delta c_{\mu-1}}&=-\frac{(2\nu-1-\Omega)
\left [(\Omega-2\mu)^2-2x_A\gamma_{AA}-1\right ]}{2x_B\gamma_{AB}
(2\mu +1-\Omega)}.
\label{coeff}
\end{align}
Substituting the coefficients in Eq.~(\ref{coeff}) into Eq.~(\ref{dl})
we find
\begin{align}
\delta l =  -\frac{|\delta c_{\mu-1}|^2}{x_B\gamma^2_{AB}
(2\mu+1-\Omega)^2}\left \{ 4x_Ax_B\gamma_{AB}^2
(2\mu-\Omega)+(2\nu-\Omega)\left [(\Omega-2\mu)^2 -
2x_A\gamma_{AA}-1\right ]^2\right \}.
\label{cons2}
\end{align}
Although it is difficult to see analytically, one can check
numerically that $\delta l$ is indeed less than zero for $\Omega
= \Omega_2$ and greater than zero for $\Omega_3$. This confirms
that $\Omega_2$ and $\Omega_3$ correspond, respectively, to the
portions of the yrast spectrum for $l < l_0$ and $l > l_0$.

\begin{center}
\begin{figure}[ht]
     \includegraphics[width=0.5\linewidth]{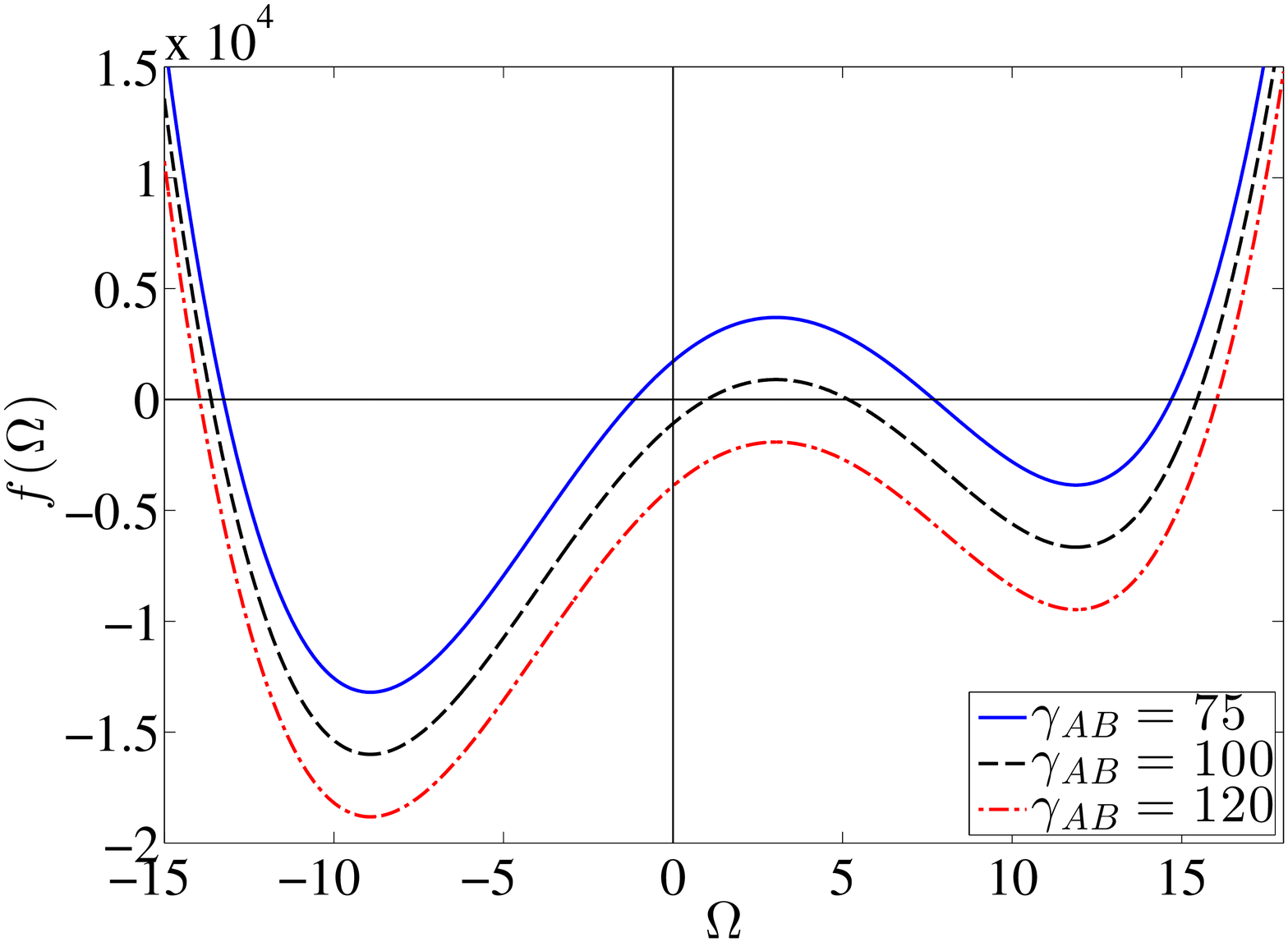}
\caption{Evolution of the function $f(\Omega)$ with variations of $\gamma_{AB}$. 
Here $\mu=0$, $\nu = 2$, $x_B = 0.2$, $\gamma_{AA}=100$ and 
$\gamma_{BB}$ = 120. The critical $\gamma_{AB}$ for which 
the second root of $f(\Omega)$ becomes zero can be obtained from
Eq.~(\ref{cons1}) and is found to be $\gamma_{AB}\simeq 91$.
The critical $\gamma_{AB}$ for which  
the double root of $f(\Omega)$ emerges can be obtained 
from Eq.~(\ref{yrcrit}) and is found to be $\gamma_{AB}\simeq 107$.} 
\label{fOmega}
\end{figure}
\end{center}

We now suppose that a gradual variation of a system parameter
leads to a violation of the inequality 
in Eq.~(\ref{stab2}). At this point, the state
$(\phi_\mu,\phi_\nu)$ ceases to be a local minimum of the energy
functional in Eq.~(\ref{Efunc}). This evolution is
easiest to visualize by considering variations in
$\gamma_{AB}$ (see Fig.~\ref{fOmega}). As $\gamma_{AB}$ is increased, 
$f(\Omega)$ shifts
down, with the effect that $\Omega_2$ increases and $\Omega_3$ 
decreases. Since $f'(0) > 0$, $\Omega_2$ will first go to zero
at some critical value of $\gamma_{AB}$, at which point $\bar
E_0(l)$ ceases to have a local minimum. This signals the fact that
persistent currents are no longer stable at $l_0$. This is consistent
 with the criterion established earlier
in Eqs.~(\ref{stab1}) and (\ref{stab2}), since by setting $\Omega=0$
in Eq.~(\ref{cons1}) one recovers precisely 
the equality corresponding to Eq.~(\ref{stab2}).
However, since $\Omega_3 > 0$ when $\Omega_2 = 0$, the
derivative discontinuity in $\bar E_0(l)$ persists, as shown schematically in
Fig.~\ref{yrast}(b). As $\gamma_{AB}$ is increased further, 
the difference between the roots $\Omega_2$ and $\Omega_3$
gradually decreases and they eventually merge into a
double root. At this point the
discontinuity at $l_0$ vanishes, as indicated schematically in Fig.~\ref{yrast}(c). In going from
Fig.~\ref{yrast}(b) to Fig.~\ref{yrast}(c) we can envisage two possible scenarios. In the first,
the plane-wave state is always an yrast state and the merging of the two roots establishes the 
critical condition for $(\phi_\mu,\phi_\nu)$ to be an yrast state. In other words, the plane-wave state at $l_0$ ceases 
to be an yrast state when the slope discontinuity vanishes. However, there is the possibility that a soliton state with an
energy lower than that of the plane-wave state at $l_0$ may appear before the merging of the double root. If this happens,
the emergence
of the soliton state defines the critical condition for the plane-wave state to be an yrast state.  In this case, the merging of the 
double root at best provides a bound on this critical condition. Whether or not this latter scenario actually occurs would have to be checked by explicit
numerical solutions of the coupled Gross-Pitaevskii equations for the condensate wave functions.

In the following, we will assume that the first scenario discussed above is valid
and therefore we will focus on the critical condition for which the quartic $f(\Omega)$ has a
double root. This occurs when the discriminant $\Delta$ of the quartic is zero, 
that is
\beq
\Delta(x_s,\gamma_{ss'},\mu,\nu) \equiv {\rm det}(S) = 0.
\label{yrcrit}
\eeq
Here the discriminant is defined by the determinant of the
so-called Sylvester matrix~\cite{Gelfand}
\beq
S=\left (\begin{array}{cccccccc}
a_4 &a_3  &a_2&a_1 &a_0&0&0 \\
0& a_4 &a_3  &a_2&a_1&a_0&0  \\
0&0 &a_4 &a_3&a_2&a_1&a_0\\
4a_4&3a_3&2a_2&a_1&0&0&0\\
0 &4a_4&3a_3&2a_2&a_1&0&0\\
0&0&4a_4&3a_3&2a_2&a_1&0\\
0&0&0&4a_4&3a_3&2a_2&a_1,
\end{array} \right ),
\eeq
where $a_4,...,a_0$ are the the coefficients of the
quartic $f(\Omega)$ in Eq.~(\ref{cons1}), namely
\begin{align}
a_4 &= 1, \nn \\
a_3 &= -4(\mu+\nu),\nn \\
a_2 &= 4(\mu^2 + \nu^2 + 4\mu\nu)-2(x_A\gamma_{AA}
+x_B\gamma_{BB}+1),\nn \\
a_1 &= -4\nu(4\mu^2-2x_A\gamma_{AA}-1)
-4\mu(4\nu^2-2x_B\gamma_{BB}-1), \nn \\
a_0 &= (4\mu^2-2x_A\gamma_{AA}-1)(4\nu^2-
2x_B\gamma_{BB}-1)-4x_Ax_B\gamma_{AB}^2.
\label{acoeff}
\end{align}

Alternatively, we can make use of the properties of $f(\Omega)$ to determine
the critical condition for which the $\Omega_2$  and $\Omega_3$ roots merge and take the 
common value $\Omega_0$. We observe that $f'(\Omega)$ is a cubic and that
$f'(\Omega) = 0$ has three roots, as can be seen from Fig.~\ref{fOmega}.
The frequency $\Omega_0$ is the root for which $f''(\Omega_0) < 0$. The condition
$f'(\Omega_0) = 0$ gives the equation
\beq
(\Omega_0 - 2\mu)\left[ (\Omega_0-2\nu)^2 -2x_B\gamma_{BB} - 1 \right ]
+(\Omega_0 - 2\nu)\left[ (\Omega_0-2\mu)^2 -2x_A\gamma_{AA} - 1 \right ]= 0.
\label{gprime}
\eeq
Furthermore, we see that $f(\Omega_0) = 0$ when
\beq
\left [ (\Omega_0-2\mu)^2 -2x_A\gamma_{AA} - 1 \right ]\left [ (\Omega_0-2\nu)^2 -2x_B\gamma_{BB} - 1 \right ]
=4x_Ax_B\gamma_{AB}^2.
\label{fOmega_0}
\eeq
When Eqs.~(\ref{gprime}) and (\ref{fOmega_0}) are used in Eq.~(\ref{cons2}), we find that 
$\delta l $ becomes zero when $\Omega = \Omega_0$. Thus the merging of the $\Omega_2$ and 
$\Omega_3$ roots can be determined by requiring that $f(\Omega) = 0$ and $\delta l =0$ be satisfied
simultaneously. These two equations, a quartic and quintic respectively, have to be solved numerically and the critical condition
found is identical to that obtained from Eq.~(\ref{yrcrit}).

Finally, we note that if the parameters $x_s$ and $\gamma_{ss'}$ 
 satisfy Eq.~(\ref{yrcrit}) they also satisfy the equations 
 \beq
\Delta(x_s,\gamma_{ss'},-\mu,-\nu)  = 0,
\label{inverivr}
\eeq
and
\beq
\Delta(x_s,\gamma_{ss'},\mu+n,\nu+n)  = 0,
\label{tranivr}
\eeq
where $n$ 
is an integer. This is due to the fact that the existence
of a double root of $f(\Omega)$ is not affected by 
inversion of the function $f(\Omega)$ with respect to the $y$ axis or
translation along the $x$ axis. As a result the critical condition 
can simply be written as
\beq
\Delta(x_s,\gamma_{ss'},0,|k|)  = 0,
\label{yrcrit1}
\eeq
where $k = \nu -\mu$. This agrees with the general observation we made
at the beginning of this section that the condition for $(\phi_\mu,\phi_\nu)$
to be an yrast state should only depend on the absolute difference of
$\mu$ and $\nu$. 
Furthermore, from Fig.~\ref{fOmega} we see that the necessary condition 
for a slope discontinuity to occur in the yrast spectrum is that Eq.~(\ref{cons1})
has four real roots. The latter is ensured if the discriminant is positive, 
namely,
\beq
\Delta(x_s,\gamma_{ss'},0,|k|) > 0.
\label{yrnc}
\eeq
Within the first scenario discussed earlier, the inequality in Eq.~(\ref{yrnc}) constitutes the condition for
$( \phi_\mu,\phi_{\nu})$ being an yrast state.

\subsection{Numerical results and discussion}
In general, the discriminant $\Delta(x_s,\gamma_{ss'},|k|)$ is a rather complex 
function of $x_B$, $\gamma_{ss'}$ and $k$ (we restrict ourselves to non-negative $k$ from now on). An exception 
occurs for $k=0$, where one finds that the inequality in Eq.~(\ref{yrnc}) simplifies to
\beq
\left (2x_A\gamma_{AA}+1\right ) \left ( 2x_B\gamma_{BB} 
+ 1\right )- 4x_Ax_B\gamma_{AB}^2 >0.
\label{crn0}
\eeq
This in fact coincides with the condition for stability of the ground state
against phase separation (see Eq.~(\ref{energetic_stability})). As shown in Ref.~\cite{Anoshkin},
this stability condition follows from the requirement
that the Bogoliubov excitations 
all have positive energies. The inequality in Eq.~(\ref{crn0}) thus ensures that the
uniform state ($l=0$) is the ground state of the system and, by virtue of the periodicity
of $e_0(l)$, the yrast states at all integral angular momenta are plane-wave states. 

The case of $k\geq 1$, namely the condition for plane-wave 
yrast states at non-integer angular momentum, is of course
 more complex.  However, in the $x_B\rightarrow 0$ limit,
one finds that the condition in Eq.~(\ref{yrnc}) reduces to 
\beq
\gamma_{AA} > 2k(k-1). 
\label{crb0}
\eeq
This suggests that $(\phi_\mu,\phi_{\mu+k})$ is
an yrast state if the interaction strength of the majority 
component satisfies Eq.~(\ref{crb0}) and the minority
concentration is sufficiently small, regardless of
the strength of $\gamma_{BB}$ and $\gamma_{AB}$. Similarly in the 
limit that $x_B\rightarrow 1$ the condition in Eq.~(\ref{yrnc}) reduces to 
\beq
\gamma_{BB} > 2k(k-1). 
\label{crb1}
\eeq

To explore the consequences of
interaction asymmetries on the emergence of 
certain plane-wave yrast states in more detail, we use the parameterization
introduced earlier, namely,
$\gamma_{AB}=\gamma$, $\gamma_{AA}=(1+\kappa_A)
\gamma$ and $\gamma_{BB}=(1+\kappa_B)\gamma$. 
For each $k$, the critical condition 
\beq
\Delta(x_s,\gamma_{ss'},0,|k|) \equiv
\tilde \Delta(x_B,\gamma,\kappa_A,\kappa_B,|k|) = 0
\label{crit_yrast}
\eeq
then defines a hypersurface in the parameter space 
spanned by $x_B$, $\gamma$, $\kappa_A$ and $\kappa_B$. We will
be primarily interested in the critical $x_B(\gamma,k)$ curves on such a
hypersurface for fixed values of $\kappa_A$ and $\kappa_B$. We remind 
the reader that these critical curves are the analogue of the solid curves in Fig.~\ref{gammaxb} which 
define when certain plane wave states become yrast states in the symmetric model. These curves
are recovered in the limit that $\kappa_A = \kappa_B =0$.

\begin{center}
\begin{figure}[ht]
  \includegraphics[width=0.49\linewidth]{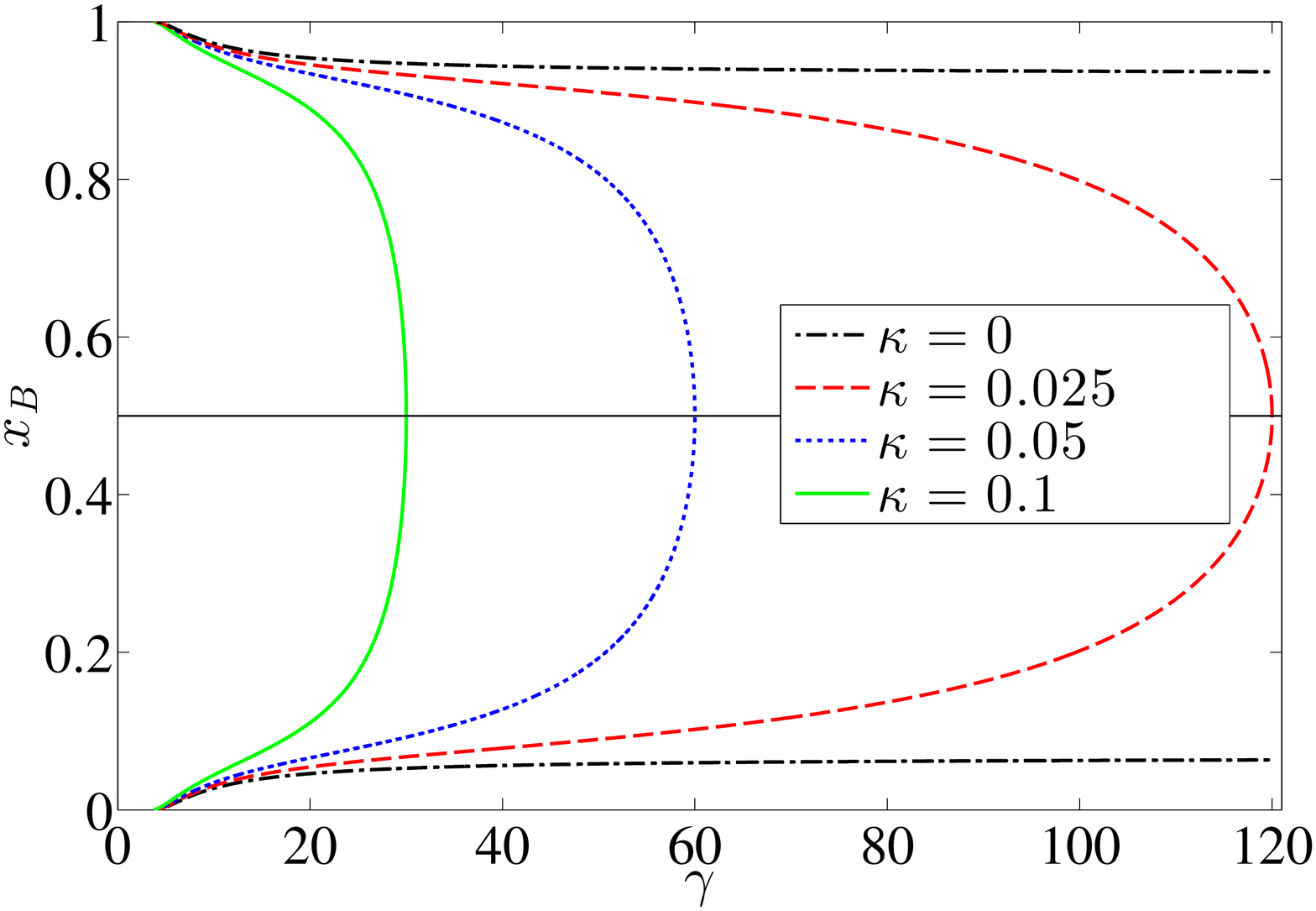}
 \includegraphics[width=0.5\linewidth]{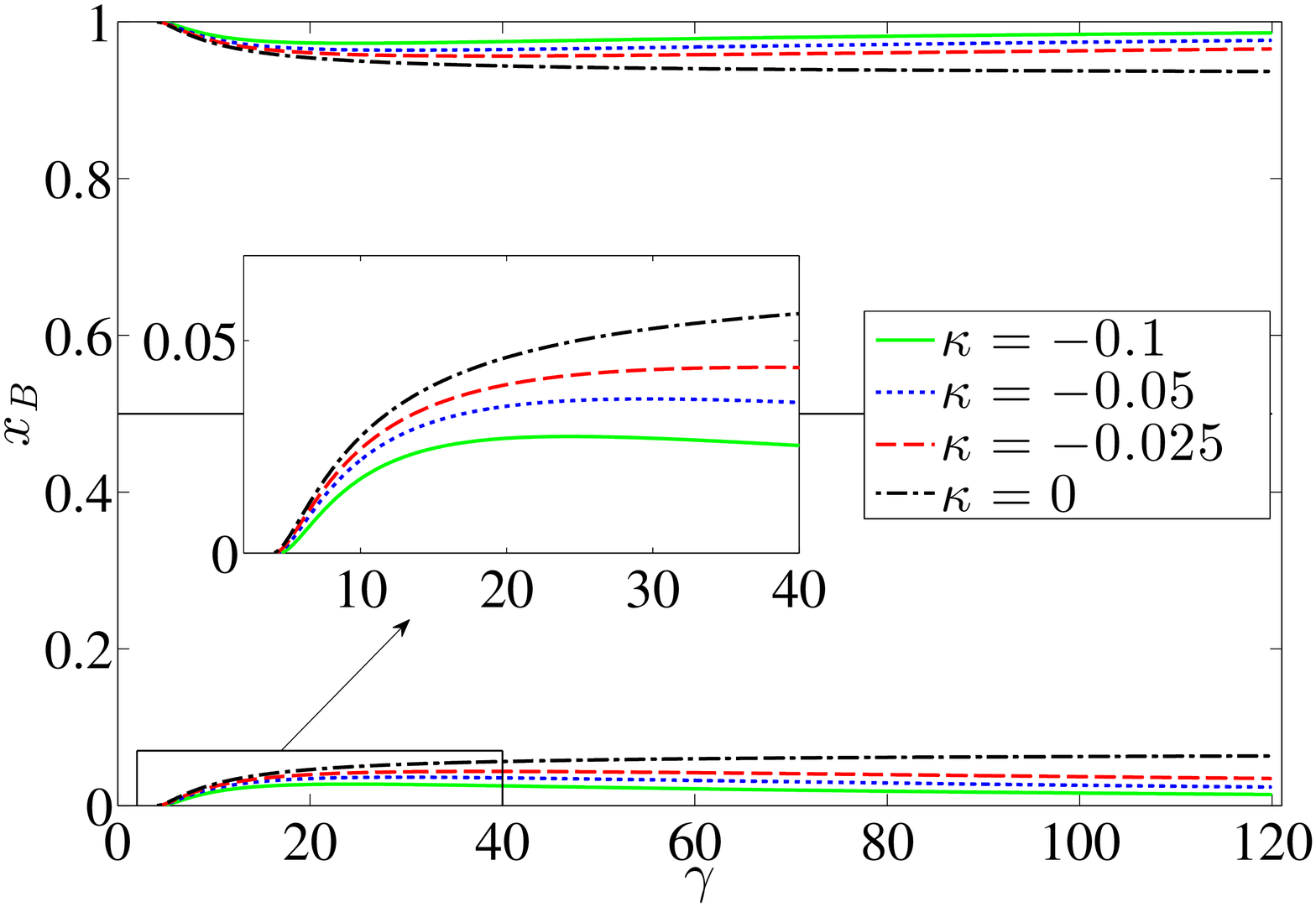}
\caption{ The plane-wave yrast state critical $x_B(\gamma,k)$ curves for $k = 2$. Here $\kappa_A=\kappa_B=\kappa$ assume different values. 
               The symmetric model is recovered in the $\kappa =0$ limit.} 
\label{xg1}
\end{figure}
\end{center}

To be specific, we first consider the case of $\gamma_{AA}=\gamma_{BB}\neq \gamma_{AB}$, 
namely $\kappa_A = \kappa_B=\kappa\neq 0$. 
In Fig.~\ref{xg1} we show the critical $x_B(\gamma,k)$ curves, determined from Eq.~(\ref{crit_yrast}), for $k=2$ and 
various values of $\kappa$. This figure shows how the limit of the symmetric model is approached as $\kappa$
tends to zero (the $k=2$ dashed curve in Fig.~\ref{gammaxb}). Our first general observation is that
these curves all possess a mirror symmetry with respect to the horizontal line $x_B= 1/2$. This is
due to the fact that for $\gamma_{AA}=\gamma_{BB}$, the discriminant $\Delta(x_s,\gamma_{ss'},\mu,\nu)$
in Eq.~(\ref{yrcrit}) is invariant under simultaneous interchanges between $x_A$ and $x_B$ and between $\mu$ and $\nu$.
As a result, we have
\beq
\tilde \Delta(x_B,\gamma,\kappa,\kappa,|k|) =  \tilde \Delta(1-x_B,\gamma,\kappa,\kappa,|k|),
\eeq 
which explains the mirror symmetry. We note that in generating Fig.~\ref{xg1}, we are no longer restricting $x_B$ to be the minority
concentration but are allowing it to vary continuously between 0 and 1. Presenting the results in this way more clearly displays
the continuous variation of the $x_B(\gamma,k)$ curves. The range $0.5\le x_B\le 1$ of course corresponds to species $A$ being 
the minority concentration and provides no new information when $\gamma_{AA} = \gamma_{BB}$. However, when $\gamma_{AA} \ne
\gamma_{BB}$, this form of the plots provides the relevant information more efficiently.

We next observe that the limiting $\kappa = 0$ curve has a horizontal asymptote for $\gamma \to \infty$~\cite{Wu}, which is approached from
one side when $\kappa <0$ and from the other when $\kappa >0$. Thus the qualitative behaviour of the $x_B(\gamma,k)$
curves is quite different in these two cases. Although all the curves have an endpoint at 
$\gamma = 2k(k-1)/(1+\kappa)$ (see Eqs.~(\ref{crb0}) and (\ref{crb1}) for $\kappa_A = \kappa_B = \kappa$)
only the curves for $\kappa >0$ (left panel in Fig.~\ref{xg1}) cross the line $x_B=1/2$ perpendicularly at 
some critical $\gamma$ value.
It can be shown from Eqs.~(\ref{gprime}) and (\ref{fOmega_0}) that this critical $\gamma$ value is given by the simple formula
\beq
\gamma_{\rm cr} = \frac{k^2-1}{\kappa}.
\eeq
If a point in the $\gamma$-$x_B$ plane lies to the right of the $x_B(\gamma,k)$ curve, then $(\phi_\mu,\phi_{\mu+k})$ is an yrast state
for the given values of the system parameters. For the example being considered in Fig.~\ref{xg1}, the point (40, 0.3) lies to the right of
the $x_B(\gamma,2)$ curve for $\kappa = 0.1$ but to the left of the curve for $\kappa = 0.05$. This
implies that the $(\phi_0,\phi_2)$ plane-wave state ceases to be an yrast state at $x_B=0.3$ as $\kappa$ is decreased 
continuously from 0.1 to 0.05. This behaviour is consistent with the discussion given in the Appendix. The value of
$\delta \bar E_{\rm int}$ given in Eq.~(\ref{deltaE_int_2}) decreases with decreasing $\kappa$ so that 
the conditions required for the $(\phi_0,\phi_2)$ plane-wave state to be an yrast state are eventually
violated. The main conclusion we reach for this kind of asymmetry is that larger positive values of $\kappa$ 
favour a plane-wave state being an yrast state.

For  $\kappa<0$ on the other hand (right panel in Fig.~\ref{xg1}), the curves are bounded by the $\kappa =0$ 
curves and the $x_B=0$ or $x_B=1$ lines and tend to these lines in the large $\gamma$ limit. We see
that the region in the $\gamma$-$x_B$ plane where the plane-wave state is an yrast state diminishes in size
as $\kappa$ is made more negative. Thus negative $\kappa$ disfavours a plane-wave state being an 
yrast state. It is clear that the conditions for a plane-wave state being an yrast state are very sensitive
to the sign of $\kappa$ and that the case of symmetric interactions ($\kappa =0$) is a very special one. The inset
to the right panel of Fig.~\ref{xg1} shows more clearly how the $x_B(\gamma,k)$ curves approach the $x_B = 0$ line 
as $\gamma \to \infty$. We have here a situation in which at some $x_B$ value, a plane-wave state can become an yrast state with
increasing $\gamma$ but then ceases to be an yrast state with further increases in $\gamma$.

\begin{center}
\begin{figure}[ht]
 \includegraphics[width=0.49\linewidth]{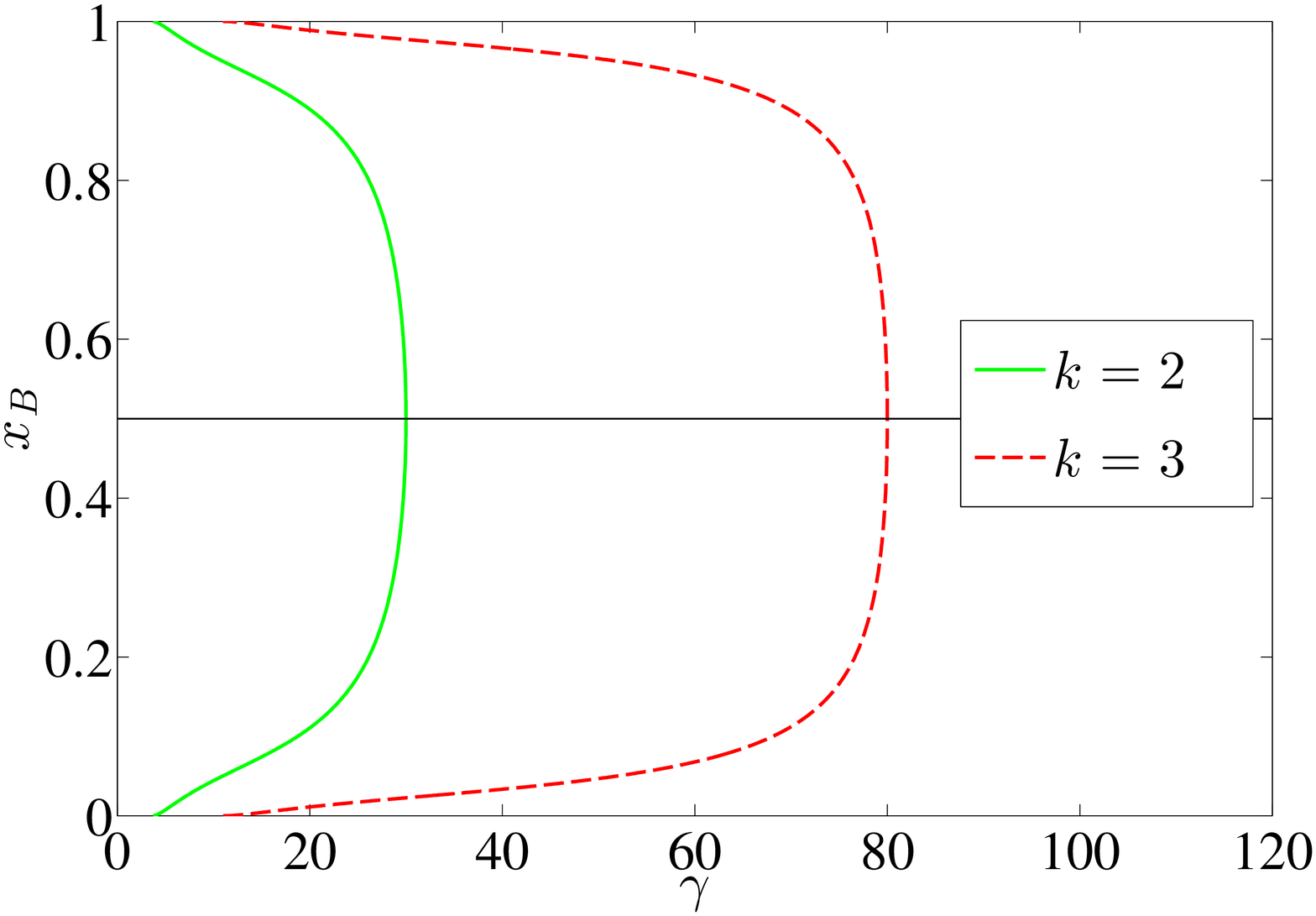} 
  \includegraphics[width=0.49\linewidth]{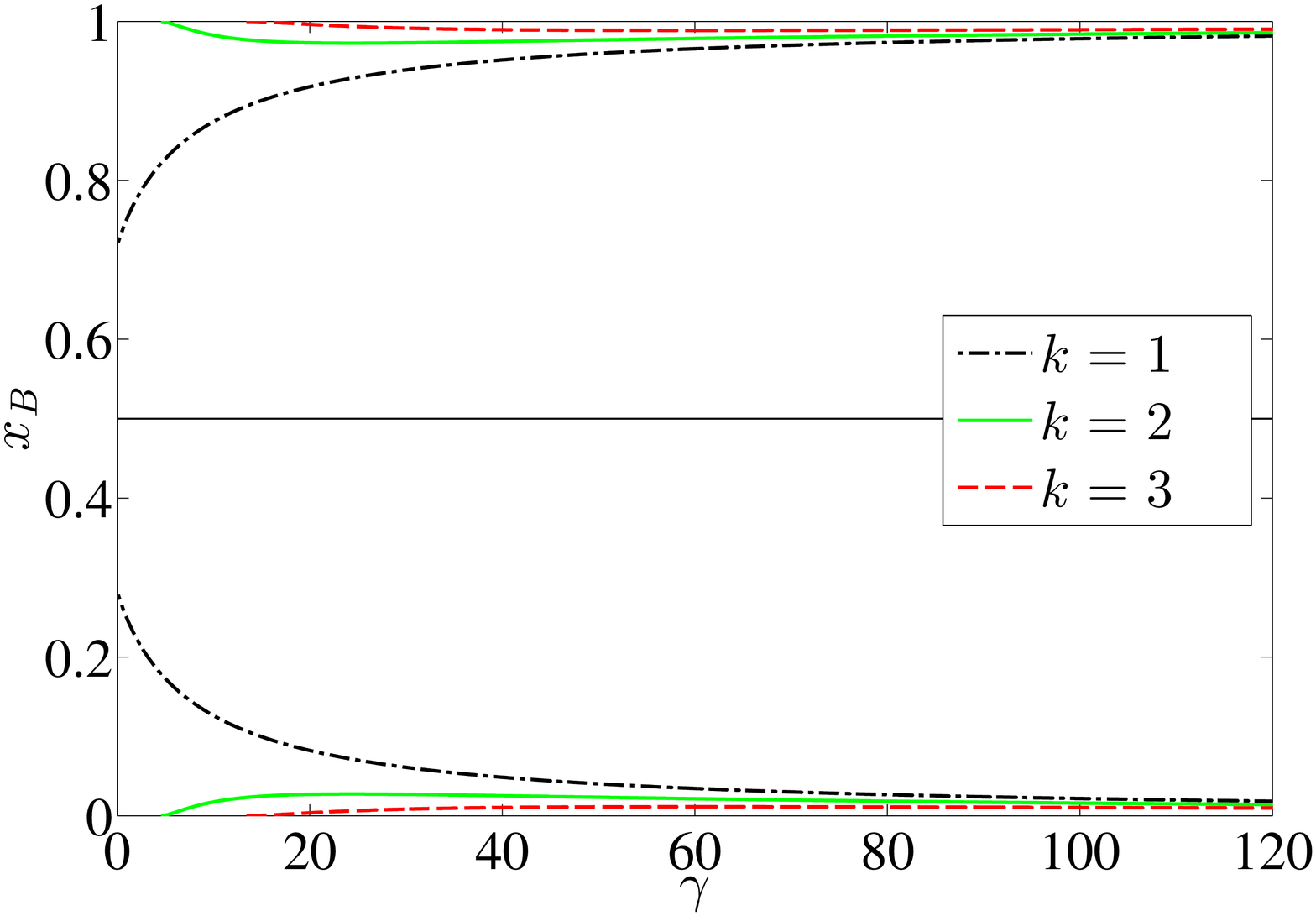} 
\caption{The plane-wave yrast state critical $x_B(\gamma,k)$ curves for $k=1,2,3$, with $\kappa= 0.1$ (left panel) and $\kappa=-0.1$ (right panel). 
As explained in the text, the plane wave states $(\phi_\mu,\phi_{\mu+1})$ are yrast states for any $x_B$ when $\kappa >0$, which explains the absence of
the $k=1$ critical curve in the left panel.} 
\label{xg2}
\end{figure}
\end{center}
In Fig.~\ref{xg2} we show the $x_B(\gamma,k)$ curves for different $k$, again for the case $\gamma_{AA}=\gamma_{BB}$.
The left panel is for $\kappa=0.1$ and the right for $\kappa=-0.1$. For the symmetric model it is known that the $(\phi_0,\phi_1)$ state
is an yrast state for all $x_B$ and any positive value of $\gamma$. As explained in the Appendix, this state is necessarily also
an yrast state when $\kappa >0$, and for this reason, there is no critical curve for $k=1$ in this case. For both signs of $\kappa$ we see
that the conditions for a plane-wave state being an yrast state become more stringent with increasing $k$. The curves are
qualitatively similar to those in Fig.~\ref{xg1} except for the $k=1$ curve with $\kappa <0$. In this case, the endpoint of the
$x_B(\gamma,1)$ curve is a point on the $x_B$-axis. As $\kappa \to 0^-$, the $x_B(\gamma,1)$ curves approach $x_B = 1/2$ and 
the $k=1$ state is an yrast state for all $x_B$ and $\gamma$. This implies that the region in the $x_B$-$\gamma$ plane {\it between}
the $k=1$ critical curves is the region where the $k=1$ state is {\it not} an yrast state. 

\begin{center}
\begin{figure}[ht]
\includegraphics[width=0.49\linewidth]{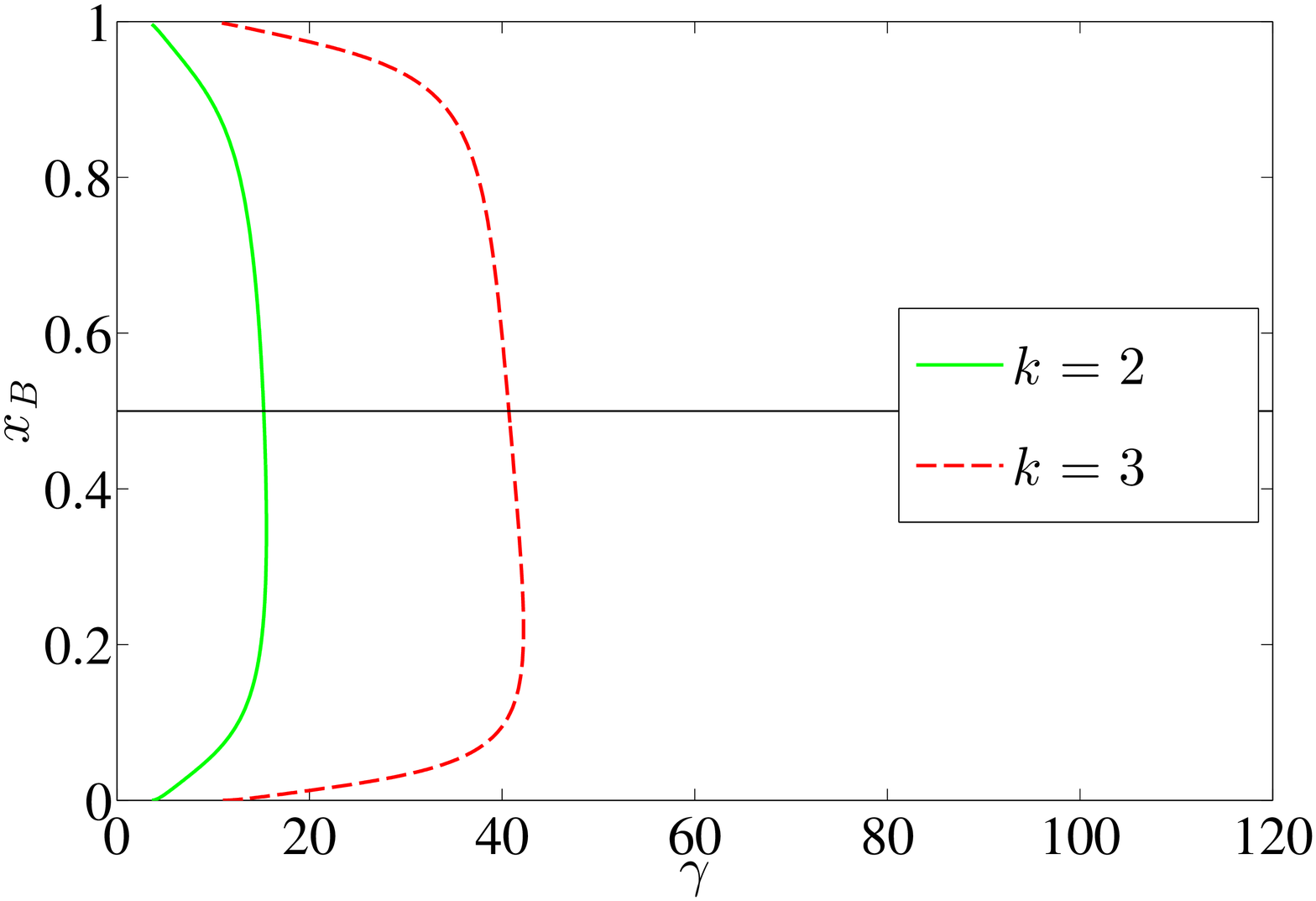}
\includegraphics[width=0.49\linewidth]{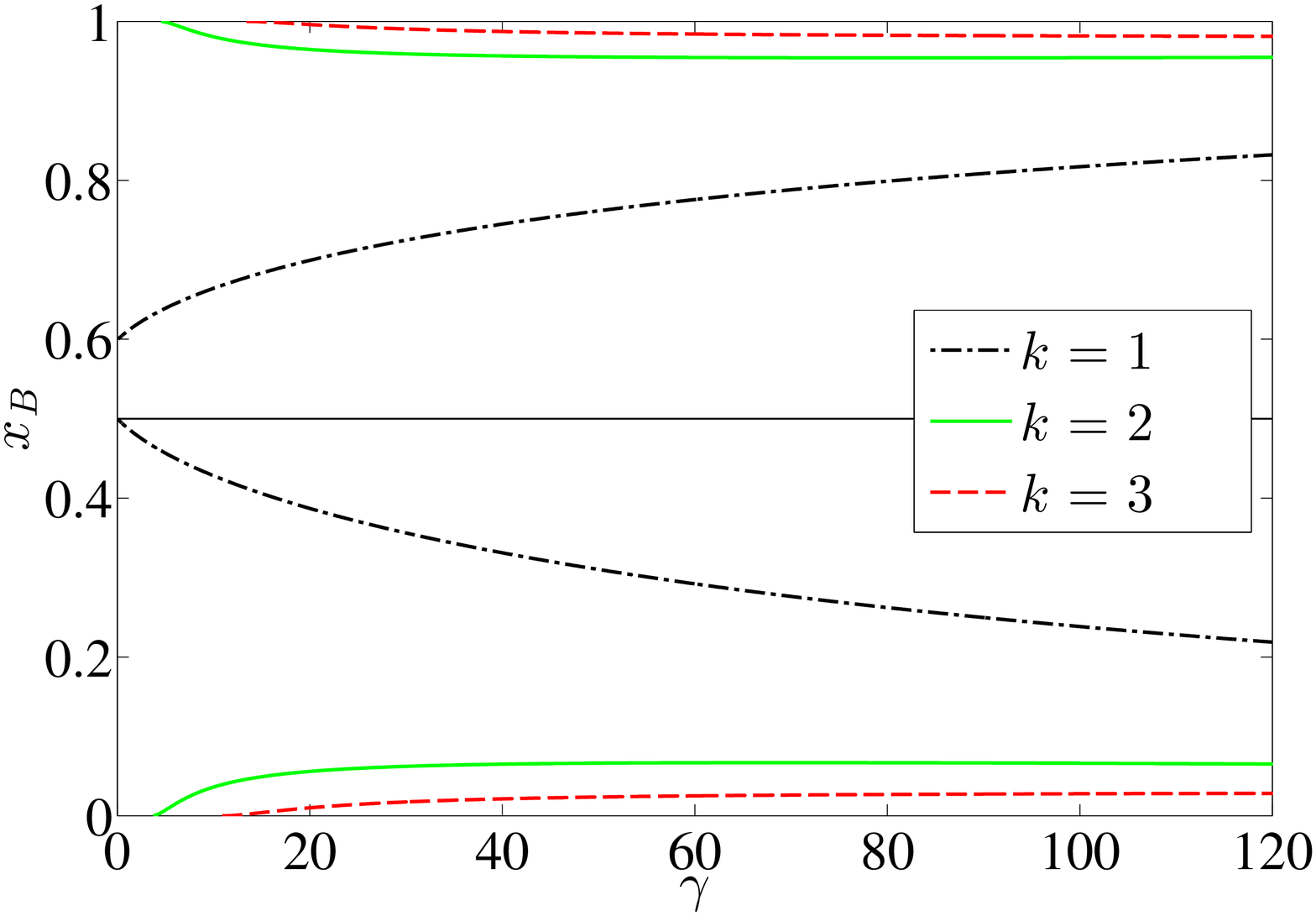} 
\caption{ The plane-wave yrast state critical $x_B(\gamma,k)$ curves for $k=1,2,3$, where $\kappa_A=0.1$ and 
$\kappa_B= 0.3$ in the left panel, and $\kappa_A=0.1$ and $\kappa_B=-0.1$ in the right panel. The $x_B(\gamma,1)$
curve is once again absent in the left panel since the $k=1$ plane-wave state is an yrast state for all $x_B$ when $\kappa_A > 0$ and 
$\kappa_B>0$.} 
\label{xg3}
\end{figure}
\end{center}
Finally we show in Fig.~\ref{xg3} some examples of critical $x_B(\gamma,k)$ curves for the system with the most general type of interaction asymmetry $\kappa_A \neq \kappa_B\neq 0$. We find that these curves are qualitatively similar to those for $\kappa_A = \kappa_B\neq 0$, with one obvious difference, namely the absence of mirror symmetry with respect to $x_B = 1/2$ line. For $0\le x_B \le 0.5$, $B$ is the minority species and $\kappa_B$ is the minority asymmetry parameter. On the other hand, for $0.5 \le x_B \le 1$, $A$ is the minority species and $\kappa_A$ is the minority asymmetry parameter. These figures can thus be viewed as providing the critical curves for two different sets of asymmetry parameters for the minority and majority species.

The curves for $k=1$ in the right panel of Fig.~\ref{xg3} show an interesting asymmetry. The $k=1$ critical curve in the range $0\le x_B \le 0.5$
has an endpoint at $x_B = 1/2$ at $\gamma = 0$. This is true whenever the minority asymmetry parameter $\kappa_B$ is less than zero. 
As $\kappa_B  \to 0^-$, this curve moves continuously to the $x_B = 1/2$ line. However, when the minority species is $A$ with $\kappa_A>0$,
the critical curve has an endpoint on the $x_B$ axis that depends on the value of $\kappa_A$. As $\kappa_A \to 0^+$,
this point moves to $x_B = 0.5$ and the whole critical curve approaches the $x_B = 1/2$ line.

\section{Concluding remarks}
In this paper we have studied the structure of the mean-field yrast 
spectrum of a two-component gas in the ring geometry with arbitrary inter-particle interaction strengths. 
In the case of the symmetric model, the nature of the spectrum can be elucidated
by means of analytic soliton solutions of the coupled GP equations~\cite{Smyrnakis2,Wu}.
Such solutions, however,  are not known for the asymmetric model in which the interaction strengths
take on different values. Nevertheless, we were able to show that some 
of the salient properties of the yrast spectrum
can be determined via a perturbative analysis of the GP energy functional. In particular, we derived criteria, expressed in terms of inequalities,
which determine whether a specific plane-wave state is a local minimum of the GP energy functional. We then assumed that the global minimum
of the energy functional on the angular momentum hypersurface corresponding to this plane-wave necessarily occurs at that particular plane-wave state.
Furthermore, if the GP energy functional
has a local minimum at this state, the yrast spectrum does as well and persistent currents are 
thus stable~\cite{Bloch} at the angular momentum of the plane-wave state. 
We then showed that the yrast spectrum at these angular momenta has slope discontinuities which persist
even after the yrast spectrum ceases to exhibit a local minimum. Finally, we showed that the plane-wave state ceases to be
an yrast state when the system parameters satisfy a certain critical condition.

In the future we plan a more detailed numerical investigation of
the yrast spectrum based on the solution of the coupled GP equations for the condensate wave functions. Such a study would provide
the solitonic portions of the yrast spectrum that join the plane-wave yrast states that we have analysed in this paper.

\acknowledgments
This work was supported by a grant from the Natural Sciences and Engineering Research Council of Canada. 
This project was implemented through the Operational Program ``Education and Lifelong Learning", Action 
Archimedes III and was co-financed by the European Union (European Social Fund) and Greek national funds 
(National Strategic Reference Framework 2007 - 2013).

\appendix
\section{}
In this Appendix, we investigate the yrast spectrum in the angular moment
interval $0\le l \le 1/2$. As discussed in Sec.~\ref{Stability}, the plane-wave states
of interest in this angular momentum range are
$(\phi_\mu,\phi_{\mu+k})$ with angular momentum $l =
\mu + kx_B$, where $k$ is an integer restricted to the range 
given by Eq.~(\ref{k-range}).
We argue that such a state can indeed be an yrast state
if the intra-species interaction strengths $\gamma_{AA}$
and $\gamma_{BB}$ are both {\it sufficiently large} in comparison 
to the inter-species interaction strength $\gamma_{AB}$. In effect 
we are claiming that conditions exist for which the state 
$(\mu, \mu + k)$ is assured to
be a global minimum on the $l=\mu + kx_B$ hypersurface. We emphasize, however, that in general
these are sufficient but not necessary conditions. As found previously, it is possible for these
plane-wave states to be yrast states even in the symmetric model where all the interaction
parameters are the same.

Our objective is to determine whether the plane-wave state $(\mu, \mu + k)$ 
can be a global minimum of the GP energy functional on the $l = \mu + kx_B$ hypersurface.
To investigate this possibility we consider the wave functions
\beq 
\psi_A = \phi_\mu+ \delta \psi_A, \quad \psi_B = \phi_\nu + 
\delta \psi_B, 
\label{tpsi}
\eeq
where $\nu = \mu + k$ and 
\beq
\delta \psi_A = \sum_m \delta c_m \phi_m ,\quad \delta \psi_B = \sum_m 
\delta d_m \phi_m.
\label{dpsi}
\eeq
Here, the deviations $\delta c_m$ and $\delta d_m$ are not
necessarily small, and for certain choices, can in fact lead to
another pair of plane waves. However, as established in Sec.~\ref{Stability}, the state $(\phi_\mu,\phi_\nu)$
has the lowest energy of all the plane wave 
states with the same angular momentum.

The difference in energy between the $(\psi_A, \psi_B)$ 
and $(\phi_\mu,\phi_\nu)$ states can be written as
\beq
\delta \bar E = \delta \bar E_K + \delta \bar E_{\rm int}.
\eeq
Here
\beq
\delta \bar E_K =x_A \sum_m m^2 |\delta c_m|^2 + x_B \sum_m m^2 
|\delta d_m|^2 +x_A \mu^2\left (\delta c_\mu + \delta c_\mu^* \right )+x_B \nu^2\left (\delta d_\nu + \delta d_\nu^* \right ) 
\label{deltaE_K0}
\eeq
and 
\beq
\delta \bar E_{\rm int} =
x_A^2\pi \gamma_{AA}\int_0^{2\pi}d\theta \left|\delta 
\rho_A(\theta)\right |^2+x_B^2\pi \gamma_{BB}\int_0^{2\pi}d\theta 
\left| \delta \rho_B(\theta)\right |^2 + 2x_Ax_B \pi 
\gamma_{AB}\int_0^{2\pi}d\theta \delta\rho_A(\theta)
\delta\rho_B(\theta),
\label{deltaE_int}
\eeq
where $\delta \rho_s = |\psi_s|^2 - |\phi_s|^2$.
We note that the change in
interaction energy $\delta \bar E_{\rm int}$ is zero whenever $(\psi_A,\psi_B)$ is a plane-wave state. 

We now consider the difference in kinetic energy $\delta \bar E_{\rm K}$
in more detail. Using the 
normalization constraints (\ref{A_norm}) and (\ref{B_norm}) in Eq.~(\ref{deltaE_K0}) we find
\beq
\delta \bar E_K =x_A \sum_m (m^2-\mu^2) |\delta c_m|^2 + x_B \sum_m (m^2 -\nu^2)
|\delta d_m|^2  
\label{deltaE_K1}
\eeq
It is apparent that the change in kinetic energy $\delta \bar E_K$ can be made negative
with appropriate wave function variations.  The argument we make depends simply on the fact that 
$\delta \bar E_K$ has a lower bound $\delta \bar E_{l.b.}$. The kinetic energy of the plane-wave state $(\phi_\mu,\phi_\nu)$ is
$\bar E_K[\phi_\mu,\phi_\nu] = x_A \mu^2 + x_B \nu^2$. Since the kinetic energy functional $\bar E_K[\psi_A,\psi_B]$ is 
positive semi-definite, the lowest possible value it can have is 0. Thus the lower bound is given by $\delta \bar E_{l.b.}
= -(x_A \mu^2 + x_B\nu^2)$. It should be noted that this lower bound is reached only for the $(\phi_0,\phi_0)$ state which does 
not lie on the angular momentum hypersurface of interest. Nevertheless, this lower bound must still be valid when
variations of the wave functions are constrained to have the desired angular momentum.

The interaction energy in Eq.~(\ref{deltaE_int}) can be written as
\beq
\delta \bar E_{\rm int} =
x_A^2\pi (\gamma_{AA}-\gamma_{AB})\int_0^{2\pi}d\theta \left|\delta 
\rho_A(\theta)\right |^2+x_B^2\pi (\gamma_{BB}-\gamma_{AB})\int_0^{2\pi}d\theta 
\left| \delta \rho_B(\theta)\right |^2 + \pi 
\gamma_{AB}\int_0^{2\pi}d\theta |\delta\rho(\theta)|^2,
\label{deltaE_int_2}
\eeq
where $\delta \rho = x_A\delta \rho_A+x_B\delta\rho_B$ is the total change in particle
density. Eq.~(\ref{deltaE_int_2}) reduces to the change in interaction energy in the symmetric
model with $\gamma = \gamma_{AB}$ when $\gamma_{AA}=\gamma_{BB} =\gamma_{AB}$.
If the $(\mu, \nu)$ plane-wave state is an yrast state in the symmetric model for this value of $x_B$ and
interaction parameter $\gamma$, then this state remains an yrast state in the asymmetric model
with $\gamma_{AA}> \gamma_{AB}$ and $\gamma_{BB}> \gamma_{AB}$ since the interaction
energy only increases while the change in kinetic energy is unaltered. If this state is not an yrast 
state in the symmetric model it is still possible that it becomes an yrast state in the asymmetric 
model. We now turn to the demonstration of this possibility.

Equation~(\ref{deltaE_int_2}) implies
\beq
\delta \bar E_{\rm int} \geq 
x_A^2\pi (\gamma_{AA}-\gamma_{AB})\int_0^{2\pi}d\theta 
\left|\delta \rho_A(\theta)\right |^2+x_B^2\pi 
(\gamma_{BB}-\gamma_{AB})\int_0^{2\pi}d\theta \left| \delta 
\rho_B(\theta)\right |^2.
\label{deltaE_int_bound}
\eeq
which is positive-definite if both
$\gamma_{AA}$ and $\gamma_{BB}$ are greater than $\gamma_{AB}$.
Since $\delta \bar E_K$ can be negative,
the question of interest is whether $\delta \bar E$ can be made
positive-definite with a suitable choice of parameters. Specifically, we 
wish to determine whether conditions exist such that
\beq
\delta \bar E_{\rm int} > |\delta \bar E_K|
\label{inequality}
\eeq
for {\it any} wave function having the same angular momentum but with a 
{\it lower} kinetic energy, i.e. $\delta \bar E_K < 0$.
\begin{center} 
\begin{figure}[ht]
     \includegraphics[width=0.49\linewidth]{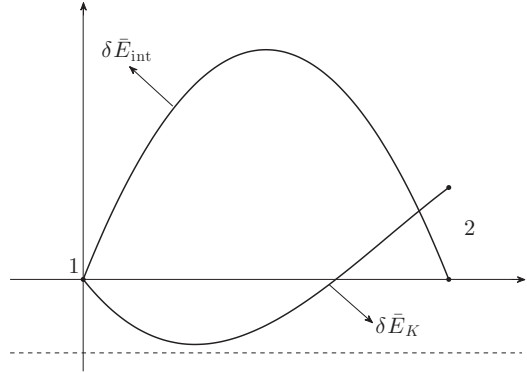}
\caption{The schematic variation of $\delta \bar E_{\rm int}$ and $\delta \bar E_K$
along a path on the angular momentum hypersurface between the kinetic energy 
minimizing plane-wave state at 1 and some other plane-wave state at 2.  $\delta \bar E_{\rm int}$ increases in magnitude
as $\gamma_{AA}$ and $\gamma_{BB}$ are increased relative to $\gamma_{AB}$. It is assumed that $\delta \bar E_K$ becomes 
negative along parts of the path; the dashed curve indicates the lower bound on $\delta \bar E_K$.} 
\label{illus}
\end{figure}
\end{center}

In Fig.~\ref{illus} we illustrate the expected qualitative variation of $\delta \bar E_{\rm int}$ and
$\delta \bar E_K$ along some path on the angular momentum hypersurface between
the plane-wave state minimizing the kinetic energy and some other plane-wave state
that has a higher kinetic energy. Along this path $\delta \bar E_{\rm int}$ is positive and 
vanishes at the ends of the path. The dashed line indicates the lower bound on 
$\delta \bar E_K$. Since $\delta \bar E_{\rm int}$ can be made arbitrarily large
by increasing $\gamma_{AA}$ and $\gamma_{BB}$ relative to 
$\gamma_{AB}$, it is clear that $\delta \bar E_{\rm int}$ can be made to satisfy the
inequality in Eq.~(\ref{inequality}) except possibly at the start of the path at 1 where it
goes to zero. However, at this point we know that, if Eqs.~(\ref{stab1}) and (\ref{stab2})
are satisfied, $\bar E$ has a local minimum at this point. Thus, even if $\delta \bar E_K$
were to decrease as one moved away from 1, the local minimum at this point would ensure
that Eq.~(\ref{inequality}) is satisfied. Since the inequalities in Eqs.~(\ref{stab1}) and (\ref{stab2}) become
even stronger with increasing  $\gamma_{AA}$ and $\gamma_{BB}$, it is clear that
it is always possible to ensure that Eq.~(\ref{inequality}) is satisfied at all points along the path from 1 to 2
where $\delta \bar E_K$ is negative. This observation implies that the plane-wave state at 1 can be made
a global minimum on the angular momentum hypersurface for a suitable choice of the interaction parameters.
In the body of the paper we make the stronger assumption that the plane-wave state is a global
minimum when the inequalities in Eqs.~(\ref{stab1}) and (\ref{stab2}) are satisfied. All our results 
are consistent with this assumption.

 \end{document}